\documentclass{aa}

\usepackage{graphicx}

\def\be{\begin{equation}}
\def\fe{\end{equation}}
\def\bea{\begin{eqnarray}}
\def\fea{\end{eqnarray}}

\def\mylabel#1{\label{#1}}

\def\O{{\cal O}}
\def\text#1{{\rm #1}}

\def\mb{\mbox}
\def\qaq{{\quad\text{and}\quad}}

\def\Lagr{{\cal L}}
\def\energy{{\cal E}}
\def\Om{\Omega}
\def\C{{\cal C}}
\def\Lie{{\cal L}}

\def\Kentr{{\cal K}}
\def\entr{\alpha}

\def\a{{i}}   
\def\b{{j}}

\def\stat{{(0)}}	

\def\X{{\!{\it X}}}	
\def\Y{{\!{\it Y}}}
\def\Z{{\!{\it Z}}}
\def\A{{\!{\it A}}}
\def\B{{\!{\it B}}}

\def\n{{\rm n}}		
\def\p{{\rm p}}		

\def\eps{\Omega}
\def\veps{{\underline{\eps}}}	

\def\S{{\cal S}}      

\def\D{{\cal D}}

\def\OmK{\Omega_{\rm K}}
\def\OmKA{\Omega_{{\rm K},{\it A}}}

\begin{document}

\title{Slowly rotating superfluid Newtonian neutron star model with
entrainment} 
\titlerunning{Slowly rotating superfluid Newtonian neutron star model ...}

\author{R. Prix\inst{1} \and  G. L. Comer\inst{2} \and 
N. Andersson\inst{1}}

\authorrunning{R.~Prix et al.}

\institute{Department of Mathematics, University of Southampton, 
Southampton SO17 1BJ, United Kingdom \and Department of Physics, 
Saint Louis University, P.O. Box 56907, St. Louis, MO 63156-0907, USA}

\date{}

\abstract{ We develop a 
formalism that can be used to model slowly rotating 
superfluid Newtonian neutron stars.  A simple two-fluid model is 
used to describe the matter, where one fluid consists of the 
superfluid neutrons that are believed to exist in the inner crust 
and core of mature neutron stars, while the other fluid is a 
charge neutral 
conglomerate of the remaining constituents (crust nuclei, 
core superconducting protons, electrons, etc).  We include  
the entrainment effect, which is a non-dissipative interaction 
between the two fluids whereby a momentum induced in one of the 
fluids will cause part of the mass of the other fluid to be carried 
along.  The equations that describe rotational equilibria (i.e. 
axisymmetric and stationary configurations) are approximated using 
the slow-rotation approximation; an expansion in terms of 
the rotation rates of the two fluids where only terms 
up to second-order are kept.  Our formalism allows the
neutrons to rotate at a rate different from that of the 
charged constituents.  For a particular equation of state 
that is quadratic in the two mass-densities and relative
velocities of the fluids, we find
an analytic solution to the slow-rotation equations.  This result
provides an elegant generalisation to the 
two-fluid problem of the standard expressions for
the  one-fluid polytrope ${\cal E} \propto \rho^2$.
The model equation of state includes entrainment and is general enough to 
allow for realistic values for, say, mass and radius of the star.  It 
also includes a mixed term in the mass densities that can be related 
to ``symmetry energy'' terms that appear in more realistic equations 
of state.  We use the analytic solution to explore how relative 
rotation between the two fluids, the ``symmetry energy'' term, and 
entrainment affect the neutron star's local distribution of particles, as
well as global quantities as
the Kepler limit, ellipticity, and moments of inertia.     
}

\maketitle

\section{Introduction}

Pulsars were first discovered over 34 years ago (Hewish et al \cite{Hetal68}).  
Since then, nearly 1300 such rapidly rotating, highly magnetised neutrons 
stars have been found, and as pointed out by Lorimer in his recent Living 
Reviews in Relativity (Lorimer \cite{L01}), 700 of these were found in the last 
4 years alone.  Without doubt, the overwhelming majority of these pulsars 
are old and cold (with core temperatures below $10^9$~K).
According to equation of state calculations,  
their interiors will contain superfluid neutrons, superconducting protons, 
a plasma of highly degenerate and ultra-relativistic electrons, and 
perhaps other more exotic particles (pions, hyperons, etc.) deep in their 
cores.  In recent years, there has been a continued effort to model 
superfluid dynamics in neutron stars in both the Newtonian (Epstein \cite{E88}; 
Mendell \& Lindblom \cite{ML91}; Mendell \cite{M91a}, \cite{M91b};
Lindblom \& Mendell  \cite{LM94}, \cite{LM95}, \cite{LM00}; Lee
\cite{ulee}; Prix \cite{rp1999}; Sedrakian \& Wasserman \cite{SW2000}; 
Andersson \& Comer \cite{AC01b}) and general relativistic regimes
(Carter \cite{C89},  Carter \& Langlois \cite{CL95a}, \cite{CL95b},
\cite{CL98}; Langlois et al \cite{LSC98}; Comer et al \cite{CLL99};  
Andersson \& Comer \cite{AC01a}).  This paper is aimed at improving
our  understanding of local and global properties of rotating Newtonian 
superfluid neutron stars.

The strongest evidence for superfluidity in the inner crust and 
core of a neutron star
(see, for example, Sauls (\cite{S89}) and references therein) is
provided by the well-known glitch 
phenomenon (the occasional sudden spin-up of relatively young 
pulsars).  Our confidence in an explanation  based on the transfer of
angular momentum between two loosely coupled components 
is bolstered by the 
fact that the neutrons and protons are described using the same many-body 
theory of Fermi liquids and BCS mechanism that has been so successful at 
describing superconductors (Pines \& Nozi\`eres \cite{PN66}).  This being the 
case, one would expect superfluidity in neutron stars to share many of
the well-established properties of laboratory 
superfluids and superconductors.  For instance, it is known that
in a mixture of two interpenetrating fluids there is a coupling that 
arises whereby the momentum of one of the liquids is not simply 
proportional to that liquid's velocity, rather, it is a linear combination 
of the velocities of both fluids.  This is the so-called entrainment 
effect and it implies that when one liquid starts to flow then it will 
necessarily induce a momentum in the other constituent.
Entrainment between protons and neutrons is a key component of models of 
neutron star superfluidity, and it is something that we will focus on 
in this study.

We develop  a formalism for describing 
slowly rotating Newtonian superfluid neutron stars which allows
the neutrons and protons to rotate at different rates.
Our analysis is based  
on a Newtonian model that is the non-relativistic limit of a comprehensive 
model developed by Carter, Langlois and their collaborators for the general 
relativistic regime (Carter \cite{C89}, Carter \& Langlois
\cite{CL95a}, \cite{CL95b}, \cite{CL98}; Langlois et al \cite{LSC98}).
Our model is simplified in the sense that it describes a superfluid
neutron star in terms of only two fluids (we refer the reader to, for
example, Comer et al (\cite{CLL99}) for justification).  One fluid  
is composed of the superfluid neutrons, existing in the inner crust and 
core, and the other fluid is a charge neutral
conglomerate of the remaining  
constituents (i.e. crust nuclei, core protons, and the crust and core 
electrons) that we will loosely refer to as the ``protons.''  

A further simplification of our model regards the vortices
of the superfluid. Because the density of neutron vortices is expected 
to be about $10^2-10^5 {\rm cm}^{-2}$  for typical pulsar rotation
rates,  it is reasonable to adopt a smooth averaged
description of vorticity on a macroscopic scale.
The resultant model then resembles a perfect fluid. 
There is, however, one important difference that arises
because of the possible interaction between the vortex lattice
and the second fluid. This is a dissipative effect usually
referred to as ``mutual friction''.  In general, this mechanism 
tends to drive the two fluids towards co-rotation, but
there are two extreme cases where a stationary description of a
two--fluid star with different rotation rates would still be
possible. The first case is (obviously) when the interaction is very
small (the ``free'' vortex limit). Somewhat surprisingly, similar results
apply  in the case when the
interaction is very strong 
(resembling the ``pinned'' vortex case in a solid
crust, cf.~Langlois et al.~\cite{LSC98}).
Although we could, in principle, allow for this second case in the present
context (as has been done in an earlier study by Prix~\cite{rp1999}),
we will refrain from doing so and assume that vortex friction
is negligible on the time-scales we are considering. This is done
because the emphasis of the present work is on the detailed role of the 
entrainment effect. Including
vortex friction would unnecessarily complicate the discussion and
distract the attention away from the new piece of physics we have 
incorporated.

We will generalize the Chandrasekhar-Milne slow-rotation method (Chandrasekhar 
\cite{Chandra33}; Milne \cite{Milne23}) to our simple two-fluid model.  That is, 
we make an expansion 
in terms of the fluid rotation rates keeping only terms up to 
second order.  We include 
entrainment, and consider the most general case where the neutrons do not 
corotate with the protons.  This relative rotation of 
neutrons to protons is also present in the general relativistic slow 
rotation scheme of Andersson \& Comer (\cite{AC01a}) (who  provide an 
extended discussion as to why such a generalization is necessary), and 
the Newtonian slow rotation scheme of Prix (\cite{rp1999}).  However, although the 
formalism developed by Andersson \& Comer does allow for entrainment, 
their  numerical study did not include it, and Prix excluded 
entrainment altogether.  As we wish to explore the effects of entrainment 
on rotational equilibria, and the current best model for entrainment is 
for the Newtonian regime (Borumand et al \cite{BJK96}), it is natural
for us to take up where Prix (\cite{rp1999}) left off, and consider
Newtonian models.  

As an application of the slow-rotation formalism, we will consider a 
particular form of the equation of state  (EOS) that is the most 
general quadratic form in the mass-densities and in the relative
velocity.  
This EOS contains a mixed term in the neutron and proton 
mass densities that can be related to so-called ``symmetry energy'' terms 
in realistic equations of state (i.e. terms that vanish when there are 
equal numbers of protons and neutrons).  Remarkably, we find an analytic 
solution to the slow-rotation equations for this EOS. 
This solution has enough free parameters that realistic neutron star values
for the mass, say, can be obtained.  Using this solution we will show
how relative rotation, entrainment, and the  
``symmetry energy'' term affect the neutron star's local distribution of 
particles, as well as global quantities like 
the ellipticity, the Kepler limit, and the moments of inertia.  

Before concluding this introduction we should note
one of the main motivations behind the present work.
Once we have developed a framework for obtaining stationary superfluid 
models, we want to study 
their dynamics as manifested by the various modes of pulsation.  This 
issue is of particular interest as it is known that several different 
modes of a rotating neutron star are generically unstable due to the 
emission of gravitational radiation (Friedman \& Schutz \cite{FS78};
Andersson \& Kokkotas \cite{NA01}). One can even plausibly speculate
that pulsation modes unique to superfluids will be excited during a
pulsar glitch and that the resultant gravitational radiation may be
detected by a future generation of advanced detectors (Andersson \&
Comer \cite{AC01c}). 
Our aim is to investigate these various possibilities by 
using the models developed here as 
background for a linear perturbation calculation of the 
relevant pulsation modes of a rotating superfluid neutron star.  

\section{Canonical two-fluid hydrodynamics}

In this section we present the Newtonian canonical description of a 
non-dissipative interacting two-fluid system, which can be derived either 
as the non-relativistic limit  (Andersson \& Comer
\cite{AC01b}) of the canonical covariant description
developed by Carter and Langlois (\cite{CL98}), or directly  
from an analogous Newtonian variational principle (Prix
\cite{rp2000}, \cite{rp2001}).  Since we are mainly interested in
describing a superfluid neutron star core we label the two fluids by
indices ``$\n$'' and ``$\p$'', representing the  
neutrons and protons, respectively.  The fundamental variables of our 
description are then the respective particle number densities $n_\n$ and 
$n_\p$ and the corresponding particle currents $\vec{n}_\n$ and 
$\vec{n}_\p$.  We will only consider situations where the particle 
numbers are conserved individually, so that we have 
\be
   \partial_t n_\n + \nabla_\a n_\n^\a = 0 \,, \quad \text{and} \quad
   \partial_t n_\p + \nabla_\a n_\p^\a = 0 \,,
\fe
where we sum over repeated spatial indices $\a,\b = 1,2,3$.  The 
respective velocities $\vec{v}_\n$, $\vec{v}_\p$ and the mass densities 
$\rho_\n$ and $\rho_\p$ follow from   
\be
   \vec{n}_\n = n_\n \vec{v}_\n \,, \quad
   \vec{n}_\p = n_\p \vec{v}_\p \,, \mylabel{equVelocities}
\fe
and
\be
   \rho_\n = m^\n n_\n \,, \quad
   \rho_\p = m^\p n_\p \,, \quad
   \rho = \rho_\n + \rho_\p \,, \mylabel{equDensities}
\fe
where $m^\n$ and $m^\p$ are the respective (fixed) masses per particle
of the two fluids, and $\rho$ is the total mass density.

The internal energy density or ``equation of state'' 
$\energy(n_\n, n_\p, \vec{n}_\n, \vec{n}_\p)$ of the two-fluid system must 
satisfy Galilean invariance (and isotropy, as we want to describe 
isotropic fluids), and therefore has to be of the form
\be
   \energy = \energy(n_\n, n_\p, \Delta^2) \,, \quad\text{where} \quad
   \Delta^2 \equiv \left(\vec{v}_{\n} - \vec{v}_{\p}\right)^2 \,.
\fe
This energy function defines the respective chemical potentials $\mu^\n$,
$\mu^\p$ and the ``entrainment function'' $\entr$ as the conjugate
variables to $n_\n$, $n_\p$ and $\Delta^2$, namely
\be
   d\energy = \mu^\n\,dn_\n + \mu^\p\,dn_\p + \entr \, d\Delta^2 \,,
              \mylabel{equFirstLaw}
\fe
which can be regarded as the ``first law of thermodynamics'' for this
system.  The canonical description of the two-fluid system is based on 
a convective variational principle (Prix \cite{rp2000},\cite{rp2001})
analogous to the method used by Carter (\cite{BC85}) in the general
relativistic context.  The dynamics  
of the system is therefore described by a Lagrangian density 
$\Lagr(n_\n, n_\p, \vec{n}_\n, \vec{n}_\p)$, which has the usual form of 
\mb{``$\Lagr =$ kinetic energy $-$ potential energy''}.  In the present 
context this means that
\be
   \Lagr = {1 \over 2} \rho_\n \vec{v}_\n^{\,2} + {1 \over 2} \rho_\p 
           \vec{v}_\p^{\,2} - \left(\energy +  \rho \Phi\right) \,,
           \mylabel{equLagrangian}
\fe
where $\Phi$ is the gravitational potential, which is related to the
total mass density $\rho$ by the Poisson equation  
\be
   \nabla^2 \Phi = 4\pi G \rho \,. \mylabel{equPoisson}
\fe

Variation of the Lagrangian density $\Lagr$ defines the momenta per 
(fluid) particle, $\vec{p}^{\,\n}$ and $\vec{p}^{\,\p}$, and the ``rest 
frame chemical potentials,'' $-p^\n_0$ and  $-p^\p_0$, as the conjugate 
variables to the currents $\vec{n}_\n$, $\vec{n}_\p$ and particle 
densities $n_\n$, $n_\p$, i.e. 
\be
   d\Lagr = \vec{p}^{\,\n} \cdot d\vec{n}_\n + 
\vec{p}^{\,\p} \cdot d\vec{n}_\p + 
            p^\n_0 dn_\n + p^\p_0 dn_\p\,.
\fe
Using the explicit expression (\ref{equLagrangian}) for the Lagrangian
density together with the ``first law'' (\ref{equFirstLaw}), we obtain
the following expressions for $p_0^\n$ and $p_0^\p$, 
\bea
    -p_0^\n &=& \mu^\n + \vec{v}_\n \cdot\vec{p}^{\,\n} - {1\over2} m^\n
    \vec{v}^{\,2}_\n  + m^\n \Phi \,, \nonumber \\
    -p_0^\p &=& \mu^\p + \vec{v}_\p \cdot\vec{p}^{\,\p} - {1\over2} m^\p
    \vec{v}^{\,2}_\p  + m^\p \Phi \,, \mylabel{equp0}
\fea
and the conjugate momenta $\vec{p}^{\,\n}$ and $\vec{p}^{\,\p}$ are
related to the particle currents as 
\bea
    \vec{p}^{\,\n} &=& \Kentr^{\n\n}\,\vec{n}_\n +
                       \Kentr^{\n\p}\,\vec{n}_\p \, \nonumber \\
    \vec{p}^{\,\p} &=& \Kentr^{\n\p}\,\vec{n}_\n +
                       \Kentr^{\p\p}\,\vec{n}_\p \,, 
                       \mylabel{equEntrainment}
\fea
where the symmetric ``entrainment matrix'' $\Kentr^{\X\Y}$ has the
following components:
\bea
    \Kentr^{\n\n} &=& {1\over n_\n^2}\left(m^\n n_\n - 2\entr\right) 
                      \,, \quad
    \Kentr^{\p\p} = {1\over n_\p^2}\left(m^\p n_\p - 2\entr\right) \,, 
                    \nonumber \\ \quad
    \Kentr^{\n\p} &=& {2\entr \over n_\n n_\p} \,.
                      \mylabel{equEntrMatr}
\fea

The general form of the entrainment relation (\ref{equEntrainment}),
which states that the momenta are linear combinations of the currents,
is a consequence of the Galilean invariance of the internal energy 
$\energy$.  The most important difference from the case of a single 
fluid is that the particle momentum of each of the two fluids is in 
general {\em not aligned} with its respective current.  From 
(\ref{equEntrainment}) we see that this deviation is caused by the 
off-diagonal matrix element $\Kentr^{\n\p}$ of (\ref{equEntrMatr}), which 
is proportional to the entrainment function $\entr$ and therefore 
expresses the dependence of the internal energy density $\energy$ on the 
relative velocity $\Delta$, as seen in (\ref{equFirstLaw}).  Intuitively 
this means that a particle current $\vec{n}_\n$ of the neutrons  
impinges some momentum on the protons and vice versa, due to the {\em 
interaction} between the two fluids.  In the absence of such an 
interaction, i.e. for $\entr=0$, the general entrainment relation 
(\ref{equEntrainment})  simply reduces to the usual single-fluid 
form, and we have \mb{$\vec{p}^{\,\n} = m^\n \vec{v}_\n$} and 
\mb{$\vec{p}^{\,\p} = m^\p \vec{v}_\p$}.

The equations of motion for the two fluids, obtained from a Newtonian 
convective variational principle (see Prix \cite{rp2000},
\cite{rp2001}; Andersson \& Comer \cite{AC01b}), thus have the 
following form:
\bea
    n_\n \left(\partial_t \,\vec{p}^{\,\n} - \vec{\nabla} p_0^\n\right) + 
    n_\n^\b \left( \nabla_\b\, \vec{p}^{\,\n} - \vec{\nabla} p^\n_\b
    \right) &=& 0 \,, \nonumber \\
    n_\p \left(\partial_t \, \vec{p}^{\,\p} - \vec{\nabla} p_0^\p\right) + 
    n_\p^\b \left(\nabla_\b\, \vec{p}^{\,\p} - \vec{\nabla} p^\p_\b\right) 
    &=& 0 \,. \mylabel{equGenEuler}
\mylabel{equEOM}
\fea
At this point, the two equations are ``formally'' uncoupled, but the 
interaction between the two fluids enters through relations (\ref{equp0}) 
and (\ref{equEntrainment}), which relate the {\em kinematical} quantities 
$n_\n$, $n_\p$, $\vec{n}_\n$ and $\vec{n}_\p$  to their respective 
conjugate {\em dynamical} quantities $p_0^\n$, $p_0^\p$, $\vec{p}^{\,\n}$ 
and $\vec{p}^{\,\p}$.  In fact, the fluids will be coupled even in the 
case of two ``non-interacting'' fluids, i.e. for an equation of state of 
the form \mb{$\energy = \energy_\n(n_\n) + \energy_\p(n_\p)$}.  This is 
simply because the two fluids live in the same gravitational potential, 
cf.~Prix (\cite{rp1999}).  

\section{Entrainment and effective masses} 

In the framework of condensed matter physics the description of 
interacting independent constituents, for example electrons moving through 
a ``background'' of ions, is usually formulated in terms of an 
``effective'' or ``dynamical'' mass (Ziman \cite{Ziman65}), rather than in the 
entrainment formalism presented in the previous section.  One reason for 
this might be that in the usual contexts one of the two constituents (the 
``background'') can usually be assumed to be at rest with respect to the 
observer.  It is then convenient to formulate the dynamics of the second 
constituent in a form that resembles the vacuum expressions, with the 
free particle mass being replaced by an ``effective mass'' that accounts 
for the interaction with the background.  

The general definition of the effective mass 
is based on the particle energy spectrum, 
$E(\vec{p})$, say, of particles with momentum $\vec{p}$ moving relative 
to the non-vacuum background.  The velocity $\vec{v}$ of the particles as 
a function of their momentum is then given by \mb{$\vec{v} = {\partial E 
/ \partial \vec{p}}$}, and the acceleration $\dot{\vec{v}}$ (where the 
dot represents the total time derivative) is linked to the force 
$\dot{\vec{p}}$ via a relation of the form
\be
   \dot{v}^i = m^{-1}_{ij} \dot{p}_j \,, \quad\text{where} \quad
   m^{-1}_{ij} \equiv {\partial^2 E \over \partial p_i \partial p_j} \,,
   \mylabel{equMassTensor}
\fe
which defines the effective mass tensor $m_{ij}$.  In the two-fluid 
problem we are interested in, the background can usually be assumed to be
isotropic, and therefore the particle energy spectrum will be of the
form $E = E(p^2)$, which implies that 
\be
   \vec{v} = \left(2{\partial E \over \partial p^2}\right) \, \vec{p} \,.
             \mylabel{equVP}
\fe
In this case the effective mass tensor of (\ref{equMassTensor}) can be
expressed as 
\be
   m^{-1}_{ij} = \left(2 {\partial E \over \partial p^2}\right) \delta_{ij}
   + \left(4 {\partial^2 E \over \partial p^2 \partial p^2}\right) p_i p_j 
   \,.
\fe
In the limit of small momentum, where the energy will be mainly quadratic 
in $p$, we can neglect the second term and define an {\em effective mass 
scalar} $m^*$ in the usual way.  This leads to 
\be
   {1 \over m^*} \equiv 2 {\partial E \over \partial p^2} \, \quad 
   \Longleftrightarrow \quad E(p) \approx {p^2 \over 2 m^*} + 
   \text{const} \ .
\fe

We see from (\ref{equVP}) that an equivalent form of this definition is 
\be
   \vec{p} = m^* \, \vec{v} \,, \mylabel{equEffM}
\fe
which clarifies the link between the effective mass and the more general 
formalism of entrainment used in the present work.  The key observation 
to make is that the effective mass description is formulated in the 
``rest-frame'' of the background.  If we choose the background to be the 
``neutrons,'' we could define the neutron rest-frame by 
\mb{$\vec{n}_\n = 0$}, and therefore the entrainment relation 
(\ref{equEntrainment}) in this frame reads as
\be
   \vec{p}^{\,\p} = \left(\Kentr^{\p\p}n_\p\right) \, \vec{v}_\p \,.
                    \mylabel{equEffM1}
\fe
By comparison with (\ref{equEffM}) we can relate the proton effective
mass $m^{p*}$ to the entrainment matrix by \mb{$m^{\p*} =
\Kentr^{\p\p}n_\p$}.  Using the explicit expression 
(\ref{equEntrMatr}) for $\Kentr^{\p\p}$ in terms of the entrainment
function $\entr$, we find 
%
\be
2\entr = n_\p ( m^\p - m^{\p*}) = \varepsilon \rho_\p \,,
\mylabel{equRelation1}
\fe
where we introduced the dimensionless ``entrainment coefficient''
$\varepsilon$ as\footnote{We note that Lindblom \& Mendell (\cite{LM00})
have used a slightly different parameter  to
characterize the entrainment effect. Their 
parameter  $\epsilon$ is related to our coefficient  $\varepsilon$ via
\begin{displaymath}
\epsilon = {\varepsilon \rho_\p \over \rho_\n - \varepsilon \rho}\,.
\end{displaymath}
}
\be
   \varepsilon \equiv {m^\p - m^{\p*}\over m^\p}\,.
\label{equEpsilon}
\fe
Therefore the entrainment matrix $\Kentr^{\X\Y}$ is determined
completely in terms of the effective mass of a single particle
species, for example the protons in the present formulation.
Of course, we also see (from (\ref{equRelation1}) by symmetry) that the
effective masses of the two species are not independent, because they
obviously have to satisfy   
\be
m^\n - m^{\n*} = {n_\p \over n_\n} ( m^\p - m^{\p*} )\,.
\fe

However, as pointed out by Carter (\cite{BC01}, private
communication), the above definition of the effective mass is not
quite unique because we have to specify what we mean by the
``rest-frame'' of the neutrons.  Indeed, we 
have defined it by setting $\vec{n}_\n = 0$.  However, we see from 
(\ref{equEntrainment}) that in this frame we generally have  
\mb{$\vec{p}^{\,\n}\not=0$}.  Therefore another equally viable choice of 
the ``neutron rest-frame'' would be given by setting 
\mb{$\vec{p}^{\,\n}=0$}.  This would then lead to 
\mb{$\vec{n}_\n\not=0$}.  Based on this second choice, the entrainment
relation (\ref{equEntrainment}) becomes
\be
   \vec{p}^{\,\p} = \left( \Kentr^{\p\p} - {(\Kentr^{\n\p})^2 \over
                    \Kentr^{\n\n}}\right) n_\p \vec{v}_\p \,,
\fe
leading to an alternative definition of the effective proton mass
$m^{\p\#}$, which is {\em not equivalent} to that of 
Eq.~(\ref{equEffM1}).  The corresponding relation between $\entr$ and 
$m^{\p\#}$ is
\be
   2 \entr = n_\p (m^\p - m^{\p\#}) \left\{ 1 + {n_\p \over n_\n} {(m^\p -
             m^{\p\#}) \over m^\n} \right\} \,. \mylabel{equRelation2}
\fe

How does this apparent ambiguity in the definition of the effective mass 
affect our attempt to model entrainment in superfluid neutron stars?  
Fortunately, it turns out that the difference between the two (equally 
valid) definitions becomes {\em numerically} small in two interesting 
special cases: 
\begin{enumerate}
    \item[(i)] neutron star matter: $m^\p \sim m^{\p\#}\sim m^\n$ and 
    $n_\p \ll n_\n$,
    \item[(ii)] electrons in a metal: $n_- = n_+$ and $m_- \sim 
    m_-^\#\ll m_+$ ,
\end{enumerate}
where in the second case $n_-$ and $n_+$ represent the number
densities of negative and positive charges respectively (i.e. electrons
and protons), and $m_-$ and $m_+$ the corresponding mass per particle.
In these two cases equation (\ref{equRelation2}) becomes approximately 
equal to the previous expression  (\ref{equRelation1}), and the two 
possible definitions lead to very similar results.

There exist in the nuclear physics literature a few calculations of the 
proton effective mass at neutron star densities.  The results are very 
much equation of state dependent, but one can extract some useful constraints 
on the range of values.  Chao et al (\cite{CCY72}) find, 
for the range $\rho_{\rm nuc} \leq \rho \leq 2.5 \rho_{\rm nuc}$ 
(where $\rho_{\rm nuc}$ represents
nuclear density, i.e. $\rho_{\rm nuc} = 2.7 \times 10^{14}$~g/cm$^3$), 
that $m^*_\p/m_\p 
\approx 0.6-0.5$; Sj\"oberg (\cite{S76}) finds, for $\rho_{\rm nuc} \leq \rho$, that 
$m^*_\p/m_\p \approx 0.6-0.4$; and more recently, Baldo et al (\cite{Betal92}) 
find, for the same range as Chao et al, that $m^*_\p/m_\p \approx 0.7$.  
Thus, we see that the proton effective mass apparently can range over 
the values $0.3 \leq m^*_\p/m_\p \leq 0.7$.  
The calculations also show that the effective proton mass varies slowly 
with the density. This means that one can, as a rough 
but reasonable approximation, 
assume that the effective proton mass remains constant 
throughout the core of a neutron star. 
This is, in fact, the case for the analytic solution 
that will be discussed later.

\section{Stationary two-fluid neutron star models}

Stationary two-fluid configurations are 
characterised by \mb{$\partial_t \vec{p}^{\,\n}=0$} 
and \mb{$\partial_t \vec{p}^{\,\p}=0$}, and therefore the equations of 
motion (\ref{equEOM}) take the form
\bea
    \nabla_\a \left[\mu^\n + m^\n \Phi -{1\over2} m^\n v_\n^2\right]
    + \left(p^\n_\b \nabla_\a v_\n^\b + v_\n^\b \nabla_\b p^\n_\a\right) 
    &=& 0 \,, \nonumber \\
    \nabla_\a \left[\mu^\p + m^\p \Phi -{1\over2} m^\p v_\n^2\right]
    + \left(p^\p_\b \nabla_\a v_\p^\b + v_\p^\b \nabla_\b p^\p_\a\right) 
    &=& 0 \,, \mylabel{equStationaryEOM}
\fea
where we have inserted the explicit expressions (\ref{equp0}) for $p^\n_0$ 
and $p^\p_0$.  We consider stationary configurations where both fluids are 
rotating uniformly with rotation rates $\Om_\n$ and $\Om_\p$ about a 
common axis (otherwise the system would not be stationary), 
\be
   v_\n^\a = \Om_\n \varphi^\a \,, \qaq
   v_\p^\a = \Om_\p \varphi^\a \,, \mylabel{equRotation}
\fe
where $\varphi^\a$ is the assumed axial symmetry vector with norm 
\mb{$(\varphi^\a \varphi_\a)^{1/2} = \varpi$}, and $\varpi$ is the 
distance from the rotation axis. In spherical coordinates 
\mb{$(r,\,\theta,\,\varphi)$} we, of course, have 
\mb{$\varpi = r\sin\theta$}.  As both fluids are flowing in the same 
direction $\varphi^\a$, we can write
\be
   v_\p^\a = {\Om_\p \over \Om_\n} \,v_\n^\a \,,
\fe
which allows us to express the momenta $\vec{p}^{\,\n}$ and
$\vec{p}^{\,\p}$ of (\ref{equEntrainment}) simply as
\bea
    \vec{p}^{\,\n} &=& M^\n \vec{v}_\n \,, \quad \text{with} \quad
    M^\n \equiv \Kentr^{\n\n} n_\n + \Kentr^{\n\p} n_\p {\Om_\p \over 
    \Om_\n} \,, \nonumber\\
    \vec{p}^{\,\p} &=& M^\p \vec{v}_\n \,, \quad \text{with} \quad
    M^\p \equiv \Kentr^{\p\n} n_\n + \Kentr^{\p\p} n_\p {\Om_\p \over 
    \Om_\n} \,. \mylabel{equ25}
\fea

Because of the axial symmetry, scalars like $M^\n$ and $M^\p$ cannot
depend on the angle $\varphi$, and therefore we have
\be
   v_\n^\b \nabla_\b M^\n = 0 \,, \qaq 
   v_\n^\b \nabla_\b M^\p = 0 \,. \mylabel{equ26}
\fe
Furthermore, since we are assuming uniform rotations, cf. 
(\ref{equRotation}), we can deduce the identity
\be
   v_\n^\b \nabla_\b v_{\n\a} = - {1\over 2} \nabla_\a v_\n^2 = 
   - \nabla_\a\left({\varpi^2\over2} \Om_\n^2\right) \,. \mylabel{equ27}
\fe
This means that the second parenthesis in the equations of motion 
(\ref{equStationaryEOM}) vanishes.  Namely, using (\ref{equ25}), 
(\ref{equ26}) and (\ref{equ27}), we find
\bea
    p^\n_\b \nabla_\a v_\n^\b + v_\n^\b \nabla_\b p^\n_\a &=& 
    \hspace{1.3em} M^\n \left({1\over2}\nabla_\a v_\n^2 + v_\n^\b \nabla_\b
    v_{\n\a}\right) = 0 \,, \nonumber \\
    p^\p_\b \nabla_\a v_\p^\b + v_\p^\b \nabla_\b p^\p_\a &=& 
    {\Om_\p\over\Om_\n}M^\p \left({1 \over 2}\nabla_\a v_\n^2 + v_\n^\b 
    \nabla_\b v_{\n\a}\right) = 0 \,. \nonumber
\fea
This reduces equations (\ref{equStationaryEOM}) to two first integrals of 
motion\footnote{An elegant and more generally applicable method of 
obtaining these first integrals is presented in Appendix 
\ref{secIntegrals}.}, namely  
\bea
    \mu^\n + m^\n \Phi - {\varpi^2\over2} m^\n \Om_\n^2 &=& \C^\n \,, 
    \nonumber\\
    \mu^\p + m^\p \Phi - {\varpi^2\over2} m^\p \Om_\p^2 &=& \C^\p \,,
    \mylabel{equIntegrals}
\fea
where the constants $\C^\n$ and $\C^\p$ are in general determined by the 
rotation rates $\Om_\n$ and $\Om_\p$.  These first integrals, together 
with Poisson's equation (\ref{equPoisson}), an equation of state
$\energy(n_\n,\,n_\p,\,\Delta^2)$ and appropriate boundary
conditions completely specify the solution.  Note that these first 
integrals are also the Newtonian limits of the general relativistic 
results obtained by Andersson and Comer (\cite{AC01a}).

It will be convenient in the following to write our equations in
dimensionless form by expressing all quantities in their ``natural
units.''  We choose these to be the radius $R$ of the (non-rotating)
star for lengthscales, its central density $\rho_{0}$ for densities, and
$1/\sqrt{4\pi G \rho_0}$ for timescales.  For simplicity of notation we 
will continue to use the same symbols for the dimensionless quantities, 
so Poisson's equation (\ref{equPoisson}) now reads
\be
   \nabla^2 \Phi(r,\theta) = \rho(r,\,\theta) \,, \mylabel{equII}
\fe
whereas the form of the first integrals (\ref{equIntegrals}) is unchanged.
We note that the ``natural rotation rate'' \mb{$\sqrt{4\pi G \rho_0}$}
is slightly greater than the maximal rotation rate, the Kepler limit, 
$\Om_{\rm K}$ at which mass shedding will occur.  

\begin{figure}
\resizebox{\hsize}{!}{\includegraphics[clip]{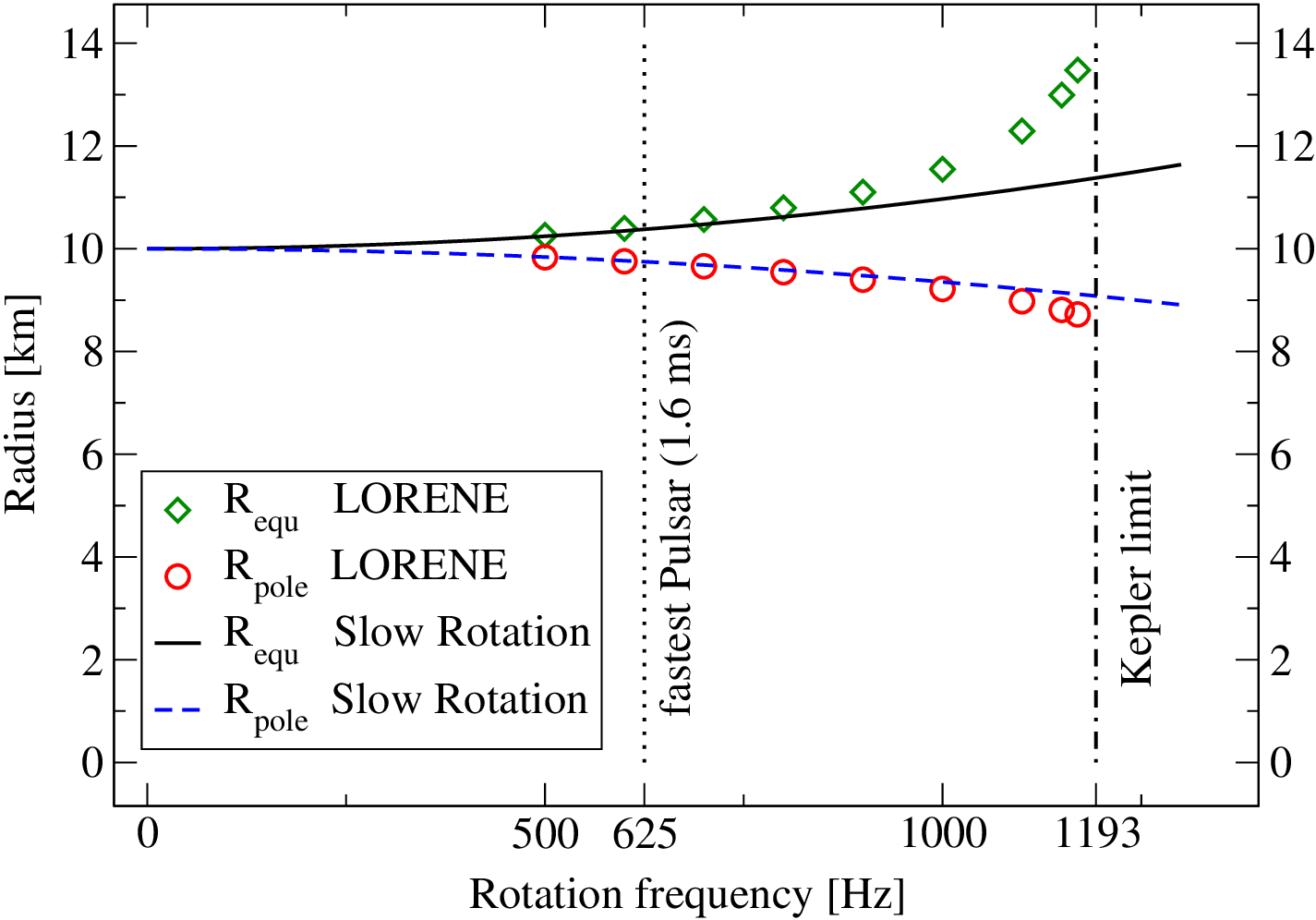}} 
\caption{Comparison of one-fluid slow-rotation configurations with 
numerically determined one-fluid configurations (obtained using the 
LORENE code). It is clear that the slow-rotation approximation leads to 
a significant underestimate of the rotational flattening 
for stars rotating near the Kepler limit.}
\label{lorene}
\end{figure}

For realistic equations of state the mass shedding limit is well 
approximated by the empirical expression \mb{$\Om_{\rm K} \approx
\sqrt{4\pi G \bar{\rho}}/3$}, where $\bar{\rho}$ is the mean density.
Therefore the dimensionless rotation rates $\Om_\n$ and $\Om_\p$ are
generally smaller than $1$, and for typical rotation rates $\Om$ of
most observed pulsars we will have 
\be
   \Om \ll 1 \,, \mylabel{equSlowRotation}
\fe
Note that the difference between the ``natural rotation rate'' and the 
mass shedding limit is further amplified by the fact that $\rho_0 \sim 
5-10 \bar{\rho}$ for a typical neutron star equation of state.  That 
being said, we must still proceed somewhat cautiously near the Kepler 
limit. A comparison between slow-rotation results and numerical 
calculations for rapidly rotating stars show that, while the slow-rotation 
approximation can accurately describe the fastest observed pulsars 
it deteriorates significantly near the Kepler limit. This is 
illustrated in Fig.~\ref{lorene}, where we compare the 
slow-rotation results for the 
equatorial and polar radii of ordinary, one-fluid stars
(governed by an equation of state of the form $\energy\propto\rho^2$)  
to accurate numerical results. The latter were obtained using the  
LORENE code, based on pseudo--spectral techniques, which 
was developed  
by the Meudon Numerical Relativity group 
(Bonazzola et al.~\cite{BGSM93}; Gourgoulhon et al.~\cite{GHLPBM99})
and which can be used to build  
rapidly rotating, Newtonian and general relativistic neutron stars.  We 
see that the slow-rotation approximation works quite well up to and 
including the fastest known pulsar, but starts to fail (by some $15\%$ to 
$20\%$) as the Kepler limit is approached.  These arguments motivate the 
range of applicability of the ``slow-rotation expansion,'' which we will 
use in the following sections.

\section{The slow-rotation approximation}


\subsection{Slow-rotation expansion of the two-fluid model}

In order to approximate the solution to (\ref{equIntegrals}) and 
(\ref{equII}), we apply a method initially due to Chandrasekhar
(\cite{Chandra33}) and Milne (\cite{Milne23}).  This method is based
on the assumption that the rotation is  
slow enough that the configuration is only slightly oblate, in such a way 
that one can express it in terms of a Taylor expansion around the 
non-rotating spherical solution.  In the following we consider such a 
``slowly-rotating'' two-fluid star. In the previous section we have seen 
that this corresponds to $\Om_\n$ and $\Om_\p$ being small compared to the
natural rotation rate $\sqrt{4\pi G\rho_0}$.  In the dimensionless 
form of the equations, the small parameters of the expansion are simply 
the rotation rates, as $\eps_\n\ll1$ and $\eps_\p\ll1$.  We can therefore 
write the solution as a Taylor expansion in $\eps_\n$ and $\eps_\p$.  In 
doing this we will neglect all terms beyond second order.

Under these assumptions any scalar physical quantity $Q$ of the rotating 
star can be written
\bea
    Q(r,\theta; \eps_\n, \eps_\p) &=& \left.Q\right|_{\eps_\n=\eps_\p=0} 
    \nonumber \\ 
    &+& \left.{\partial Q \over \partial \eps_\n}\right|_0 \eps_\n + 
    \left.{\partial Q \over \partial \eps_\p}\right|_0 \eps_\p \nonumber \\
    &+& {1\over2}\left.{\partial^2 Q \over \partial \eps_\n^2}\right|_0 
    \eps_\n^2 + \left.{\partial^2 Q \over \partial \eps_\n \partial 
    \eps_\p}\right|_0 \eps_\n\eps_\p\nonumber \\
    &+& {1 \over 2} \left.{\partial^2 Q \over \partial \eps_\p^2}\right|_0 
    \eps_\p^2 + \O(\eps^3) \,. \mylabel{equExpansion1}
\fea
Furthermore, because we are actually expanding in the ``vector''
\mb{$(\eps_\n, \eps_\p)$}, it will be convenient to introduce the
notation  $\veps$ for the ``constituent vector'' with components
$\eps_\X$, where $\X = \n,\p$ is the constituent index.  Thus we have 
\be
   \veps = \left(\eps_\n, \, \eps_\p\right) \,, \quad \text{i.e.} \quad
   \left(\veps\right)_\X = \eps_\X \,.
\fe
Terms of odd-order in $\veps$ in the expansion (\ref{equExpansion1})
will vanish identically because the configuration has to be invariant
under a simultaneous inversion of {\em both} rotation rates, i.e. for
\mb{$\eps_\n \rightarrow -\eps_\n$} and \mb{$\eps_\p \rightarrow -\eps_\p$}.  

We know that the static solution $Q|_{\veps=0}$ is purely spherical and 
denote it $Q^\stat(r)$, i.e.  
\be
   Q(r,\theta; \veps=0) = Q^\stat(r)\,.
\fe
With these definitions, the second-order expansion (\ref{equExpansion1}) 
of $Q$ can be written in a more compact form as 
\be
   Q(r,\theta; \veps) = Q^\stat(r) + \delta Q(r,\theta) + \O(\veps^4) \,
   \mylabel{equExpansion2}
\fe
with
\be
   \delta Q = \eps_\X Q^{\X\Y} \eps_\Y \,,
\fe
where we automatically sum over repeated constituent indices
(this convention is assumed from now on), and
$Q^{\X\Y}(r,\theta)$ is the symmetric expansion coefficient matrix  
\be
   Q^{\X\Y}(r,\theta) \equiv \left.{\partial^2 Q \over \partial \eps_\X
   \partial \eps_\Y} \right|_{\veps=0} \,.
\fe

It has been stated several times in the literature (Monaghan \& Roxburgh 
\cite{MR65}; Hartle \cite{Hartle67}; Smith \cite{Smith75},
\cite{Smith76}; Tassoul \cite{Tassoul78}) that the slow-rotation  
expansion of the density $\rho(r,\theta)$ breaks down when approaching the 
surface of the non-rotating star, i.e. for \mb{$r\rightarrow 1$}, because 
there \mb{$\rho^\stat(1)=0$}.  However, it is evident from 
(\ref{equExpansion2}) that the validity of
this expansion is completely independent of the values of $Q^\stat(r)$.  
The only condition for its validity is that the neglected terms 
$\O(\veps^4)$ are effectively small compared to the $\O(\veps^2)$ terms 
included in the analysis.

We now expand all physical quantities of (\ref{equIntegrals}) and 
(\ref{equII}) up to second-order in $\veps$:
\bea
    \Phi(r,\theta) &=& \Phi^\stat + \delta\Phi \,, \qquad 
    \delta\Phi = \eps_\X \,\Phi^{\X\Y}\, \eps_\Y \,, \mylabel{equExpPhi}\\
    \rho(r,\,\theta) &=& \rho^\stat + \;\delta \rho \,, \qquad 
    \;\delta\rho = \eps_\X \,\rho^{\X\Y}\, \eps_\Y \,, \\ 
    n_\A(r,\theta) &=& n_\A^\stat \;+ \delta n_\A \,, \qquad 
    \delta n_\A = \eps_\X \,n_\A^{\X\Y}\,\eps_\Y \,, \mylabel{equExpn} \\ 
    \mu^\A(r,\theta) &=& \mu^{\A\stat} + \delta\mu^\A \,, \qquad
    \delta\mu^\A = \eps_\X \,\mu^{\A,\,\X\Y}\,\eps_\Y \,, 
                   \mylabel{equExpmu} \\
    \C^\A  &=& \C^{\A\stat} +  \;\delta\C^\A \,,\qquad
    \delta\C^\A = \eps_\X \,\C^{\A,\,\X\Y} \,\eps_\Y \,. \mylabel{equExpC}
\fea
The total mass density $\rho$ is, of course, linked to the individual
particle densities by (\ref{equDensities}), i.e. \mb{$\rho = m^\A\,n_\A$}.  
Therefore we have the relations 
\be
   \rho^\stat = m^\A\, n_\A^\stat\,,\qaq
   \rho^{\X\Y} = m^\A \, n_\A^{\X\Y}\,.
\fe
Inserting the various expansions into (\ref{equIntegrals}) and 
(\ref{equII}), the zeroth-order equations, which determine the 
non-rotating configuration, are found to be
\bea
    \mu^{\A\stat} + m^\A \Phi^\stat &=& \C^{\A\stat} \,, 
    \mylabel{equStatic1} \\
    \nabla^2 \Phi^\stat(r) &=& m^\A n_\A^\stat(r) \,, \mylabel{equStatic2}
\fea
while the second-order coefficients of (\ref{equExpPhi})-(\ref{equExpC}) 
satisfy 
\bea
    \mu^{\A,\,\X\Y} + m^\A \Phi^{\X\Y} - {\varpi^2\over2}
    \delta^{\A,\,\X\Y} &=& \C^{\A,\,\X\Y} \,, \mylabel{equI2} \\
    \nabla^2 \Phi^{\X\Y} &=& m^\A n_\A^{\X\Y} \,. \mylabel{equII2}
\fea
In equation (\ref{equI2}) we have introduced the constant matrices
$\delta^{\A,\,\X\Y}$, which are defined as
\be
   \delta^{\n,\,\X\Y} \equiv \left(\begin{array}{rr} m^\n & 0 \\
   0 & 0 \end{array}\right) \,, \quad
   \delta^{\p,\,\X\Y} \equiv \left(\begin{array}{rr}
   0 & 0 \\
   0 & m^\p \end{array}\right) \,,
\mylabel{equConstdelta}
\fe
These allow us to write the second-order terms $m^\n\eps_\n^2$ and 
$m^\p\eps_\p^2$ in (\ref{equIntegrals}) as
\be
   m^\n\eps_\n^2 = \eps_\X \,\delta^{\n,\,\X\Y} \, \eps_\Y \,
\fe
and
\be
   m^\p\eps_\p^2 = \eps_\X \,\delta^{\p,\,\X\Y} \, \eps_\Y \,.
\fe

\subsection{Reduction to a single equation}

We can find an algebraic relation between the chemical potential
coefficients $\mu^{\A,\,\X\Y}$ and the density coefficients $n_\A^{\X\Y}$, 
which will allow us to considerably simplify the problem.  This relation 
is found by expanding the chemical potential $\mu^\A(n_\n, n_\p, 
\Delta^2)$ defined in (\ref{equFirstLaw}) in its arguments up to 
second-order in $\veps$.  Using (\ref{equExpn}) we find
\bea
    \mu^\A(n_\n,\, n_\p,\, \Delta^2) &=& \mu^\A(n_\n^\stat, n_\p^\stat, 0) 
    \nonumber\\
    &+& \eps_\X \, \left(\left.{\partial \mu^\A \over \partial 
    n_\Z}\right|_0 n_\Z^{\X\Y} \right. \nonumber \\
    &+& \left.\left.{\partial \mu^\A \over \partial 
    \Delta^2}\right|_0 \varpi^2 \Delta^{\X\Y}\right) \eps_\Y + 
    \O(\veps^4) \,, \mylabel{equMuExpansion}
\fea
where we have defined the constant matrix $\Delta^{\X\Y}$ as 
\be
   \Delta^{\X\Y} \equiv \left(\begin{array}{rr} 1 & -1 \\
   - 1 & 1 \end{array} \right) \,, \mylabel{equConstDelta}
\fe
This definition allows us to write the relative velocity squared, namely 
\mb{$\Delta^2=\varpi^2 \left(\eps_\n - \eps_\p\right)^2$}, in the form
\be
   \Delta^2 = \varpi^2 \,\eps_\X\,\Delta^{\X\Y} \, \eps_\Y \,.
\fe

The partial derivatives in (\ref{equMuExpansion}) are evaluated for the
non-rotating configuration, and thus they are assumed to be {\em known} 
functions, depending only on the static solution for a given equation of 
state \mb{$\energy(n_\n,\,n_\p,\,\Delta^2)$}.  Thus we define the 
``density structure function'' $\S_{\X\Y}(r)$ as
\be
   \S_{\X\Y}(r)\equiv \left(\left.{\partial \mu^\X \over \partial 
   n_\Y}\right|_0\right)^{-1} = \left(\left. { \partial^2 \energy \over 
   \partial n_\X \partial n_\Y }\right|_0 \right)^{-1} \,,
   \mylabel{equDensityStruct}
\fe
which is symmetric (and we assume invertible with inverse
$\left(\S^{-1}\right)^{XY}$), and the ``entrainment
structure function'' $\beta^\X(r)$ as
\be
   \beta^\X(r) \equiv \left.{\partial \mu^\X \over \partial 
   \Delta^2}\right|_0 = \left.{\partial^2 \energy \over \partial n_\X 
   \partial \Delta^2} \right|_0 = \left.{\partial \entr \over \partial 
   n_\X } \right|_0\,,
\mylabel{equEntrStruct}
\fe
where we used the definition (\ref{equFirstLaw}) of the chemical 
potentials $\mu^\X$ and the entrainment function $\entr$.  Comparing 
(\ref{equMuExpansion}) to the expansion (\ref{equExpmu}) we can identify
\bea
    \mu^{\A\stat} &=& \mu^\A(n_\n^\stat, n_\p^\stat, 0)\,, \\
    \mu^{\A,\,\X\Y} &=& \left(\S^{-1}\right)^{\A\B} \,n_\B^{\X\Y} + 
    \varpi^2 \beta^\A \Delta^{\X\Y}\,, 
\fea
and we  have thus arrived at an algebraic relation between the 
$\mu^{\A,\,\X\Y}$ and the $n_\B^{\X\Y}$.

Inserting this relation into the first integral (\ref{equI2}), we obtain 
an explicit expression for the density coefficients $n_\A^{\X\Y}$ in terms 
of the gravitational potential coefficients $\Phi^{\X\Y}$.  Specifically, 
we have
\bea
   n_\A^{\X\Y} &=& \S_{\A\B} \left[ C^{\B,\,\X\Y} + {\varpi^2 \over 2}
   \left(\delta^{\B,\,\X\Y} - 2 \beta^\B \Delta^{\X\Y}\right)\right. 
   \nonumber \\
   &-&\left.m^\B \Phi^{\X\Y} \right] \,. \mylabel{equNMu}
\fea
Introducing the ``derived'' background functions
\bea
    E_\A^{\X\Y}(r) &\equiv& {1 \over 3} \S_{\A\B}(r) \left(
    \delta^{\B,\,\X\Y} - 2 \beta^\B(r) \Delta^{\X\Y}\right) \,, \nonumber\\
    k_\A(r) &\equiv& \S_{\A\B}(r) m^\B \,, \mylabel{equDerivedStruct}
\fea
equation (\ref{equNMu}) can be written
\be
   n_\A^{\X\Y} = \S_{\A\B}(r) \C^{\B,\,\X\Y} + {3 \varpi^2 \over 2} 
   E_\A^{\X\Y}(r) - k_\A(r) \,\Phi^{\X\Y} \,, \mylabel{equNMu2}
\fe
which is a reformulation of the first integrals of motion (\ref{equI2})
in terms of the second--order coefficients.  

The total mass density coefficient $\rho^{\X\Y}$ can now be written 
\bea
   \rho^{\X\Y} &=& m^\A n_\A^{\X\Y} \nonumber \\
               &=& m^\A \S_{\A\B}(r) \C^{\B,\,\X\Y} + {3\varpi^2 \over 2} 
                   E^{\X\Y}(r) - k(r) \Phi^{\X\Y} \,, \mylabel{equRhoXY}
\fea
where we have introduced the further abbreviations
\bea
    E^{\X\Y}(r) &\equiv& m^\A E_\A^{\X\Y}(r) \nonumber \\ 
                &=& {1\over3} m^\A \S_{\A\B}(r) \left(\delta^{\B,\,\X\Y} 
                    - 2\beta^\B(r)\,\Delta^{\X\Y} \right) \, 
                    \mylabel{equDerivedStruct2}\\
    k(r) &\equiv& m^\A k_\A(r) = m^\A \S_{\A\B}(r) \,m^\B\,, \nonumber
\fea
which are ``known'' functions of the non-rotating star's configuration, 
determined in terms of the ``structure functions'' $\S_{\X\Y}(r)$ and 
$\beta^\X(r)$, together with the constants $\delta^{\A,\,\X\Y}$ and 
$\Delta^{\X\Y}$ and the constants of integration $\C^{\A,\,\X\Y}$.  These 
constants are determined by the boundary conditions.  Inserting 
(\ref{equRhoXY}) into Poisson's equation (\ref{equI2}) reduces the problem 
to a single partial differential equation for the coefficients 
$\Phi^{\X\Y}(r,\theta)$, namely 
\be
   \nabla^2 \Phi^{\X\Y} + k(r) \Phi^{\X\Y} = {3\varpi^2\over2}
   E^{\X\Y}(r) + m^\A \S_{\A\B}(r) \C^{\B,\,\X\Y} \,. \mylabel{equPDE}
\fe
Given the non-rotating background, a solution to this equation fully 
specifies a slowly-rotating, two-fluid configuration. 

We note that a necessary condition for this method to work, i.e. for
(\ref{equPDE}) to be well-defined, is that the structure function
$\S_{\X\Y}(r)$ is regular everywhere inside the star.  However, it is 
well-known that in the case of a single fluid the standard polytropic 
equations of state, i.e.~\mb{$\energy=\kappa n^{1+1/N}$}, leads either to 
a vanishing (for $N>1$) or an infinite density gradient (for $N<1$) at the 
surface.  This behaviour could make (\ref{equDensityStruct}) singular at 
the surface.  In order to clarify this point, we can express the definition
(\ref{equDensityStruct}) equivalently as
\be
   {n_\X^{\stat}}'(r) = \S_{\X\Y} \, {\mu^{\Y\stat}}'(r)\,, \mylabel{equ61}
\fe
where a prime denotes the radial derivative $d/dr$.
It is seen from (\ref{equStatic1}) that the chemical potentials
$\mu^{\X\stat}(r)$ and their derivatives are regular everywhere, even at
the surface.  Therefore we can conclude that the structure function
$\S_{\X\Y}$ behaves like the density gradients \mb{${n_\X^\stat}'$}
at the surface.  The slow rotation expansion used in the present work is 
therefore well-defined for stars with finite or zero density gradients at 
the surface, but is not applicable to stars with an infinite density 
gradient at the surface, similar to the $N<1$ case for the polytropes, as 
the structure function $S_{\X\Y}$ then diverges at the surface.

Inserting (\ref{equStatic1}) into (\ref{equ61}), we obtain 
expressions for the derived structure functions $k_\A(r)$
(\ref{equDerivedStruct}) and $k(r)$, which indicate their
physical meaning.  Using (\ref{equDerivedStruct}) we have
\be
   {n_\A^\stat}' = - k_\A \, {\Phi^\stat}' \quad
   \Longrightarrow\quad k_\A = - {d\n_\A^\stat \over d\Phi^\stat}\,,
\label{equkA}
\fe
and from the definition (\ref{equDerivedStruct2}) it follows that
\be
   {\rho^\stat}' = - k \,{\Phi^\stat}' \quad
   \Longrightarrow\quad k = - {d\rho^\stat \over d\Phi^\stat} \,. 
   \mylabel{equk}
\fe
From these relations we see that the ``structure'' functions 
 $k_\A(r)$
reflect the change of the individual number 
densities with respect to the gravitational
potential, while the function  $k(r)$ reflects the corresponding 
change in the total mass density. These relations  proved very useful 
in deriving the results discussed in Section~7. 

\subsection{Separation of variables}

So far, the slow-rotation approximation has allowed us to reduce the 
initial problem (\ref{equIntegrals}) and (\ref{equII}) to a single 
equation (\ref{equPDE}) for a single unknown function 
$\Phi^{\X\Y}(r,\theta)$.  We now separate the variables $r$ and $\theta$ 
by expanding $\Phi^{\X\Y}$ in the orthogonal basis of Legendre polynomials 
$P_l(\theta)$.  These are the eigenfunctions of the operator 
$\nabla^2_\theta$, satisfying 
\be
   \nabla^2 P_l(\theta) = -{l (l + 1) \over r^2} P_l(\theta) \,.
\fe
Therefore, writing 
\be
   \Phi^{\X\Y}(r,\,\theta) = \sum_{l=0}^\infty \Phi^{\X\Y}_l(r) 
        \,P_l(\theta) \mylabel{equLegendrePhi}
\fe
leads to
\be
   \nabla^2 \Phi^{\X\Y} = \sum_{l=0}^\infty P_l(\theta) \, \D_l 
        \Phi^{\X\Y}_l(r) \,, \mylabel{equLegendreEqu}
\fe
where the differential operator $\D_l$ is defined as
\be
   \D_l \equiv  {d^2 \over dr^2} + {2 \over r} {d \over dr} - {l (l + 1) 
        \over r^2}\,,
\fe
which represents the radial part of the Laplacian with an additional
``angular momentum'' part proportional to \mb{$l(l + 1)$}.

Using (\ref{equLegendreEqu}) together with the identity
\be
   {3 \varpi^2 \over 2} = r^2 \left[1 - P_2(\theta)\right] \,,
\fe
we arrive at a series of ordinary differential equations for the 
coefficients $\Phi^{\X\Y}_l(r)$ in (\ref{equLegendrePhi}).  We have
\bea
    \D_0 \Phi^{\X\Y}_0 + k\,\Phi^{\X\Y}_0 &=& r^2\,E^{\X\Y}
           + m^\A \S_{\A\B}\C^{\B,\,\X\Y}, \mylabel{equODE0} \\
    \D_2 \Phi^{\X\Y}_2 + k\,\Phi^{\X\Y}_2 &=& - r^2 
           \, E^{\X\Y} \,, \mylabel{equODE2} \\
    \D_l \Phi^{\X\Y}_l + k \, \Phi^{\X\Y}_l &=& 0 \,,
\quad \text{for} \quad l \ge 4\mylabel{equODEhom}
\fea
where only terms of even $l$ will be non-vanishing because of the expected 
equatorial symmetry of the solution.  Furthermore, we will see in the 
following section that the boundary conditions are {\em homogeneous}, and 
therefore the homogeneous equations (\ref{equODEhom}) have only trivial 
solutions, i.e. 
\be
   l \ge 4: \quad \Phi^{\X\Y}_l(r) = 0 \,.
\fe
The only non--vanishing contributions to $\Phi^{\X\Y}(r,\,\theta)$ 
therefore consist of the two functions $\Phi^{\X\Y}_0(r)$ and 
$\Phi^{\X\Y}_2(r)$, and the problem has been reduced to that of solving 
the two differential matrix equations (\ref{equODE0}) and
(\ref{equODE2}).  This is, of course, in complete analogy with the
standard result of the single fluid slow-rotation approximation. 

Once the solutions $\Phi^{\X\Y}_l$ to the differential equations
(\ref{equODE0}) and (\ref{equODE2}) have been obtained for a given 
equation of state (subject to the boundary conditions to be discussed in 
the next section), the individual density coefficients $n_A^{\X\Y}$ are 
determined from (\ref{equNMu2}).

\section{Boundary conditions and integration constants} \label{secBC}

\subsection{Boundary conditions}

As Eqs.~(\ref{equODEhom}) are linear, second-order
differential equations, two boundary conditions are necessary for each
equation.  In the case of the $l = 0$ equation (\ref{equODE0}) we need
two further conditions in order to fix the two constants of integration
$\C^{\n,\,\X\Y}$ and $\C^{\p,\,\X\Y}$.  First, we require that the 
solution must be regular at the centre of the star.  At \mb{$r=0$} the 
differential operator $\D_l$ is singular, which leads to the first 
boundary conditions 
\bea
    l = 0 &:& \quad {\Phi^{\X\Y}_0}'(0) = 0 \,, \nonumber \\
    l \ge 2&:& \quad \Phi^{\X\Y}_l(0) = 0, \, \qaq {\Phi^{\X\Y}_l}'(0) = 
                0\,, \mylabel{equBC1}
\fea
where the primes represent the radial derivative $d/dr$.  Secondly, we 
require continuity of the gravitational potential and its derivative 
across the star's surface.  That is, we have to match the interior 
solution $\Phi(r,\,\theta)$  to the exterior gravitational potential 
$\psi(r,\,\theta)$ at the surface of the rotating star $R(\theta)$:
\be
   \left.\Phi\right|_{R(\theta)} = \left.\psi\right|_{R(\theta)} \,, \qaq
   \left.\Phi'\right|_{R(\theta)} = \left.\psi'\right|_{R(\theta)} \,.  
   \mylabel{equMatching}
\fe
Here, the radial derivative deviates from the normal derivative with 
respect to the actual surface $R(\theta)$ only by terms $\O(\veps^4)$, 
which have been neglected. 

We write the slow-rotation expansion of the exterior potential $\psi$ in 
the usual way as
\be
   \psi(r,\,\theta) = \psi^\stat(r) + \delta \psi(r,\,\theta)+
   \O(\veps^4) \,,
\fe
with
\be
   \delta\psi = \eps_\X \, \psi^{\X\Y}\eps_\Y \,,
\fe
and expand the second-order coefficient in Legendre polynomials:
\be
   \psi^{\X\Y}(r,\,\theta) = \sum_{l=0}^\infty \psi^{\X\Y}_l(r) 
   P_l(\theta) \,. \mylabel{equ73}
\fe
Because $\psi$ is the solution to the Laplace equation \mb{$\nabla^2 \psi 
= 0$} (normalised by \mb{$\lim_{r\rightarrow\infty}\psi=0$}), we know the 
radial eigenfunctions for all $l$, namely
\be
   \psi^{\X\Y}_l(r) = {\kappa^{\X\Y}_l \over r^{l+1}} \,,
    \mylabel{equPsiCoeff}
\fe
where the $\kappa^{\X\Y}_l$ are constant coefficients.  The continuity 
condition (\ref{equMatching}) for the {\em static} solution implies
\be
   \Phi^\stat(1) = \psi^\stat(1) \, \qaq
   {\Phi^\stat}'(1) = {\psi^\stat}'(1) \,, \mylabel{equStat1}
\fe
and because the static potential $\Phi^\stat$ at the surface also 
satisfies the Laplace equation, i.e. \mb{$\left.\nabla^2 
\Phi^\stat\right|_{r=1}=0$}, the second derivatives must also be 
identical: 
\be
   {\Phi^\stat}''(1) = {\psi^\stat}''(1) \,. \mylabel{equStat2}
\fe

Up to second-order in $\veps$ the surface of the rotating star is 
\be
   R(\theta) = 1 + \delta R(\theta) + \O(\veps^4) \,, \mylabel{equSurface}
\fe
where we have used $R^\stat=1$.  This allows us to write the second-order 
expansion of the internal and external potentials at the matching surface 
$R(\theta)$ as
\bea
    \left.\Phi\right|_{R(\theta)} &=& \Phi^\stat(1) + {\Phi^\stat}'(1) \, 
    \delta R + \delta \Phi(1,\theta) \,, \\
    \left.\psi\right|_{R(\theta)} &=& \psi^\stat(1) + {\psi^\stat}'(1) \, 
    \delta R + \delta \psi(1,\theta) \,,
\fea
and of the radial derivatives:
\bea
    \left.\Phi'\right|_{R(\theta)} &=& {\Phi^\stat}'(1) + 
    {\Phi^\stat}''(1) \,\delta R + \delta \Phi'(1,\theta) \,, \\
    \left.\psi'\right|_{R(\theta)} &=& {\psi^\stat}'(1) + 
    {\psi^\stat}''(1) \,\delta R + \delta \psi'(1,\theta) \,.
\fea
Therefore the matching conditions (\ref{equMatching}) together with 
(\ref{equStat1}) and (\ref{equStat2}) simply reduce to
\be
   \delta \Phi(1,\,\theta) = \delta \Psi(1,\,\theta) \,, \qaq
   \delta \Phi'(1,\,\theta) = \delta \Psi'(1,\,\theta) \,,
   \mylabel{equPreBC2}
\fe
which shows that the resulting boundary condition is --- 
since there is no terms involving
$\delta R$
--- {\em independent}
of the actual second-order surface of matching.  It is thus not necessary to determine the 
perturbed surface of the star  explicitly (by the 
condition \mb{$\left.\rho\right|_{R(\theta)}=0$}).  Using (\ref{equ73}), 
(\ref{equPsiCoeff}) and (\ref{equLegendrePhi}) for the second-order terms 
$\delta\psi$ and $\delta\Phi$ in (\ref{equPreBC2}) directly results in 
the second boundary condition 
\be
   {\Phi^{\X\Y}_l}' (1) + (l+1) \,\Phi^{\X\Y}_l(1) = 0 \,, \quad
   \text{for} \quad l \ge 0 \,. \mylabel{equBC2}
\fe

\subsection{Fixing the integration constants}

As stated earlier, the $l=0$ equation requires two additional conditions 
in order to fix the constants of integration $\C^{\n,\,\X\Y}$ and 
$\C^{\p,\,\X\Y}$.  These conditions are crucial as they determine which 
type of rotating star sequence (as a function of the rotation rates 
$\Omega_\n$ and $\Omega_\p$) the solution will describe.  Among the many 
possible sequences, two seem particularly useful and will be described 
here.  The first is the {\em fixed central density} sequence (FCD), which 
is probably the most straightforward and therefore most commonly 
considered choice.  

The FCD sequence is characterized by the simple condition (for $\A = 
\n,\,\p$)
\be
    \left.n_\A\right|_{r=0} = \left. n_\A^\stat\right|_{r=0} 
    \quad \Longrightarrow \quad n^{\X\Y}_{\A,\,0}(0) = 0 \,.  
\fe
This condition obviously implies that the respective total masses will
change as functions of the rotation rates.  This sequence therefore
does not describe the same physical star at different rotation rates,
but it has the advantage of leading to a very simple condition.  From 
(\ref{equNMu}) we find that we should require 
\be
\C^{\A,\,\X\Y} = m^\A \Phi^{\X\Y}_0(0)\,. \mylabel{equFCD}
\fe

The FCD sequence is, however, not particularly 
relevant from a physical point of view.  For example, if we consider an 
isolated neutron star that is spinning down due to a magnetic torque we 
would expect the central density to increase as the star becomes less 
oblate.  The physically most interesting sequence of rotating stars
thus corresponds to requiring a {\em fixed mass} (FM). In 
other words,  one has to impose the condition of constant respective
total masses (or equivalently, total particle numbers) of the two fluids 
if the sequence is to describe the same physical star at different 
rotation rates.  This means that (for $\A = \n,\,\p$)
\be
\int n_\A(r,\,\theta) \, dV = \int n_\A^\stat(r) \, dV \,.
\fe
To second-order in the rotation rates, this condition reduces to 
\be
   \int_0^1 r^2 n_{\A,\,0}^{\X\Y}(r) \, dr = 0 \,. \mylabel{equFM}
\fe
Inserting the explicit expression (\ref{equNMu2}) for the
$n_{\A,\,0}^{\X\Y}$ component yields the integral condition (for $\A, \B 
= \n,\,\p$)
\be
   \int_0^1 r^2 \left[ \S_{\A\B} \C^{\B,\,\X\Y} + r^2 E_\A^{\X\Y} - k_\A
   \Phi_0^{\X\Y}(r) \right] \, dr = 0 \,,
\fe
which allows us to determine the constants $\C^{\A,\,\X\Y}$ for the fixed
mass sequence.  However, in practice one only needs to evaluate this 
integral for one of the two fluids, because the second fixed mass 
condition can be replaced by the equivalent but simpler requirement of 
fixed {\em total mass}, i.e. the integral over \mb{$\rho = m^\X\, n_\X$}, 
which analogously to (\ref{equFM}) leads to 
\be
   \int_0^1 r^2 m^\A n_{\A,\,0}^{\X\Y} \, dr= 0 \,.
\fe
The $l=0$ component of Poisson's equation (\ref{equII2}), i.e. 
\be
   {1 \over r^2} \left(r^2 {\Phi_0^{\X\Y}}'(r)\right)' = m^\A 
   n_{\A,\,0}^{\X\Y} \,,
\fe
allows us to reduce this condition to
\be
   {\Phi_0^{\X\Y}}'(1) = 0 \,, \quad \Longleftrightarrow \quad
   \Phi_0^{\X\Y}(1) = 0 \,,
\fe
where the second (equivalent) expression has been obtained from the
second boundary condition (\ref{equBC2}).

\section{Ellipticities, moments of inertia, and Kepler rotation rate}

Before proceeding to discuss results for a specific model equation of state
it is useful to digress on what kind of information we want to extract 
from the calculation.  There are obviously many alternative ways of 
describing rotating configurations.  We will focus our attention on the 
ellipticities, the moments of inertia and the Kepler rotation rate.
The first and second are interesting because they highlight how
rotation affects the shape and mass--distribution of the star, while
the last describes the limit of rotation at which mass-shedding at the
equator sets in.   
 
We can explicitly express the respective fluid surfaces $R_\A(\theta)$ 
to second-order in the rotation rates.  Starting from the obvious
definition  
\be
   \left. n_\A(r,\,\theta)\right|_{r=R_\A(\theta)} = 0 \,, 
\fe
and writing the second-order surfaces as
\be
   R_\A(\theta) = R_\A^\stat + \delta R_\A(\theta) + \O(\veps^4) \,,
\fe
we can express the second-order correction terms as
\be
   \delta R_\A(\theta) = \left.- {\delta n_\A(r, \theta) \over
   {n_\A^\stat}'(r)} \right|_{R_\A^\stat} \,, \label{equRadii}
\fe
where primes again denote radial derivatives.  We define the ellipticity 
of a surface in terms of the radii of the equator $R_{\rm equ}$ and of 
the pole $R_{\rm pole}$ as
\be
   e \equiv {R_{\rm equ}  - R_{\rm pole} \over R_{\rm equ}} \,.
\fe
To second-order, this means that
\be
   e_\A = {3 \over 2 R_\A^\stat} \left| {\delta n_{\A,2} \over 
          {n_\A^\stat}'}\right|_{R_\A^\stat}\,.\mylabel{equEllipticity}
\fe
Here and in the following we will assume that the two surfaces
coincide in the static case, i.e. \mb{$R_\n^\stat = R_\p^\stat = 1$},
in order to simplify the discussion. 

The respective moments of inertia 
$I_\A$ to second-order can be written as 
\be
   I_\A = I^\stat_\A + \delta I_\A + \O(\veps^4) \,,
\fe
where the second order correction term can be seen to be
\be
   \delta I_\A =  m \int_{V^\stat_\A} \varpi^2 \delta n_\A(r,\theta) \, 
                  dV\,,
\fe
which reduces to the explicit integral
\be
   \delta I_\A = {8 \pi m \over 3} \int_0^1 \left(\delta n_{\A,0}(r) -
                 {1 \over 5} \delta n_{\A,2}(r)\right) r^4 dr\,.
\label{equIA}
\fe

The Kepler rotation rate $\OmK$ is reached when one of the two fluids
spins at the rate  of a test particle in Keplerian orbit around the
equator of the star, i.e.  
at \mb{$\theta=\pi/2$} and $r=R_\A(\pi/2)$, where we assume that $\A$
represents the ``outer'' fluid at the equator.
This means that
\be
   \left(\partial_r \Phi(r,\pi/2) - r \OmK^2 \right)_{r=R_\A(\pi/2)} = 0 
   \,.
\fe
Due to the rotation--induced change in the shape of the star, the
Kepler rate also has a $\O(\veps^2)$ correction, and so we write
\be
   \OmKA^2 = \Omega_\stat^2 + \delta\OmKA^2 + \O(\veps^4) \,, 
             \label{fullkep} 
\fe
where $\Omega_\stat$ is the Kepler rate around a spherical star, i.e. 
\be
   \Omega_\stat^2 \equiv {\Phi^\stat}'(1) = {4\over3}\pi G\bar{\rho} \,.
\fe
Using the second-order expansions for $R_\A(\theta)$ and $\Phi$ this 
becomes
\be
   \delta\OmKA^2 = - 3 {\Phi^\stat}'(1) \,\delta R_\A(\pi/2) +
            \delta \Phi'(1,\pi/2) + \O(\veps^4)\,,
\label{equdKepler}
\fe
and further using (\ref{equRadii}) and (\ref{equkA}), we can
reformulate this as 
\be
   \delta \OmKA^2 = - {3 \over k_\A(1)} \delta n_\A(1,\pi/2) + \delta 
            \Phi'(1,\pi/2) \,. \mylabel{equOmK}
\fe
The Kepler rotation rate $\OmK$ determines the maximal rotation rate of 
the respective fluids before mass-shedding will occur at the equator,
i.e. if $\A$ denotes the ``outer'' fluid at the equator, then
\be
   \Omega_\A \le \OmKA\,.
\fe
Thus (\ref{equOmK}) is a quadratic expression that determines the 
respective maximal rotation rates of the two fluids.
Given the results shown in Fig.~\ref{lorene} we would expect the above 
equations 
to determine the Kepler limit to within 10-20\%.

\section{An analytic solution} \label{solution}

We have now completed the description of our slow-rotation formalism for 
two-fluid systems.  Given any suitable equation of state (which must  
provide all the relevant parameters, eg. pertaining to entrainment) 
equations (\ref{equODE0}) and (\ref{equODE2}) can be solved to produce a 
rotating configuration. 
For a typical equation of state (EOS) the 
calculation will obviously require numerics.  Although we have written a 
code that solves this problem, we will not discuss such numerical results 
here.  Instead we focus our attention on a somewhat surprising fact: It is 
possible to find an analytic solution
to our equations, including the 
general case where the two fluids are rotating at different rates, 
for a particular class of EOS. This model EOS corresponds to 
constant structure functions 
$\S_{\X\Y}$ and $\beta^\X$ (and therefore also constant coefficients $k$,
$E^{\X\Y}$...). As we will argue below, 
the resultant model is reasonably realistic and we expect it to 
prove useful in future studies of the dynamics of superfluid 
neutron stars.  Before deriving the analytic solution, we shall analyze this
particular class of EOS and assess its physical relevance.

\subsection{The ``analytic'' equation of state} \label{analeos}

Within the present slow-rotation approximation any equation of state can 
(quite generally) be expressed as
\bea
   \energy(n_\n,\, n_\p,\, \Delta^2) &=& \energy^\stat(n_\n,\, n_\p) +
   \entr^\stat(n_\n,\, n_\p)\,\Delta^2 \nonumber \\
   &+& \O(\veps^4) \,, \mylabel{equExpEOS}
\fea
since the relative velocity $\Delta$ is $\O(\veps)$.  We recall that the 
derived ``structure functions'' $E^{\X\Y}$, $k$ etc.~of 
(\ref{equDerivedStruct}) and (\ref{equDerivedStruct2}) are determined in 
terms of the two basic structure functions, the ``density structure'' 
$\S_{\X\Y}(r)$, cf.~(\ref{equDensityStruct}), and the ``entrainment 
structure'' $\beta^\X(r)$, cf. (\ref{equEntrStruct}).  In the 
slow-rotation expansion (\ref{equExpEOS}), they become
\be
   \S_{\X\Y} = \left({\partial^2 \energy^\stat \over \partial n_\X
               \partial n_\Y} \right)_0^{-1}\,,\qaq
   \beta^\X = \left({\partial \alpha^\stat \over \partial n_\X}\right)_0 
              \,. \mylabel{equ95}
\fe
The condition of constant structure functions determines the 
following class of equations of state:
\footnote{It is possible to add linear terms to $\energy^\stat$ and a
constant to $\entr^\stat$, but this does neither change the resulting
structure functions nor the static solution, and such terms have
therefore been omitted.}
\be
   \energy = {1\over2}n_\X\,A^{\X\Y} \, n_\Y + \Delta^2 B^\X n_\X \,,
   \mylabel{equAnalyticEOS}
\fe
with constant coefficients $A^{\X\Y}$ (symmetric) and $B^\X$.  Therefore
(\ref{equ95}) leads to the identifications 
\be
   \S_{\X\Y} = \left(A^{-1}\right)_{\X\Y} \,, \qaq
   \beta^\X = B^\X \,.
\fe

This class of EOS is the natural generalization to the 
two-fluid case of the ``analytic polytrope'' for single fluids 
($\energy \propto \rho^2$)  
for which the rotational corrections can be found 
analytically (Chandrasekhar and Lebovitz \cite{CL62}).  This model EOS also 
contains the simple case of the sum of two such polytropes without 
entrainment, i.e. \mb{$\energy\sim A^{\n\n} n_\n^2 + A^{\p\p} n_\p^2$}, 
for which the corresponding analytical slow-rotation solution has already 
been found by Prix (\cite{rp1999}).

The major novelty of (\ref{equAnalyticEOS}) concerns the inclusion of 
entrainment, and it is obviously important to evaluate how well this model 
approximates ``realistic entrainment.''  From the relation 
(\ref{equRelation1}) between entrainment and effective masses we see that 
the equation of state (\ref{equAnalyticEOS}) is simply characterized by a 
{\em constant effective mass} $m^{\p*}$.  We noted earlier that nuclear 
physics calculations of the effective proton mass imply that this is not an 
unreasonable approximation, even in the extreme context of 
neutron stars.  This suggests that, not only does (\ref{equAnalyticEOS}) 
have the attractive feature of leading to an analytical solution for 
slowly rotating configurations,  it also 
appears to provide a reasonable approximation which should correctly 
reflect the main qualitative features of neutron star matter (in 
particular the entrainment).  

Another novel feature of (\ref{equAnalyticEOS}) is the off-diagonal term 
$A^{\n\p}$, which leads to the term $S_{\n\p}$.  It will turn out that 
the combination   
\be
   \sigma \equiv {\S_{\n\p} \over \S_{\p\p}} 
          = {\partial \mu_\n/\partial n_\p \over \partial 
            \mu_\n/\partial n_\n} \mylabel{equ117}
\fe
plays an especially important role in determining 
the structure of our  neutron 
star model.  For a reasonable EOS, the matrix $\S_{\X\Y}$ has to be positive 
definite, which implies the constraint
\be
   \S_{\n\p}^2 < \S_{\n\n} \S_{\p\p} \,. \mylabel{equ118}
\fe
This then determines the range of acceptable values for $\sigma$.  For 
realistic situations the proton fraction will be \mb{$x_\p < 0.5$}, in 
which case one finds \mb{$\S_{\p\p} < \S_{\n\n}$}.  Therefore we can be 
sure that in this case (\ref{equ118}) is satisfied if 
\mb{$\sigma\in[-1,1]$}.  We have checked the reasonableness of this range 
by determining $\sigma$ for a more realistic EOS  due to Prakash 
et al (\cite{PAL88}).  For the range of densities and proton fractions to be 
considered here, we find that  $\sigma$ typically 
decreases monotonically from values slightly larger than one at normal 
nuclear matter density $\rho_{\rm nuc}$, to values around $- 0.4$ at about $10 
\rho_{\rm nuc}$, with the zero occurring at around $2 \rho_{\rm nuc}$.  
Thus, it seems 
that restricting a constant $\sigma$ to the range \mb{$\sigma\in[-1,1]$} 
is reasonable.  Finally, we note that $\sigma$ is intimately related 
to the so-called ``symmetry energy'' term in the Prakash et al (\cite{PAL88}) 
EOS, which is designed to vanish whenever there are equal numbers of 
neutrons and protons.  

\subsection{The static solution}

We begin by discussing the static configuration, which is the solution
to (\ref{equStatic1}) and (\ref{equStatic2}).  
The chemical potentials 
(\ref{equFirstLaw}) for the ``analytic'' equation of state
(\ref{equAnalyticEOS}) are
given by 
\be
   \mu^{\X\stat} = A^{\X\Y} n_\Y^\stat \,,
\fe
which vanish at the surface of the star, and therefore the static 
constants of integration $\C^{\X\stat}$ are determined from
(\ref{equStatic1}) at the surface $r=1$, i.e.
\be
   \C^{\X\stat}  = m^\X\, \Phi^\stat(1)\,.
\fe
Eq. (\ref{equStatic1}) can now be expressed in the form
\be
   n_\X^\stat(r) = k_\X \left[\Phi^\stat(1) - \Phi^\stat(r)\right] \,,
                   \mylabel{equ95b}
\fe
and therefore the ratio of the respective densities in the static
configuration with this  equation of state is seen to be a constant, i.e.
\be
   {n_\p^\stat(r)\over n_\n^\stat(r)} = {k_\p \over k_\n} \,.
      \mylabel{equProtonFraction}
\fe
This implies that the two fluids share a common outer surface at $r=1$.
Inserting the densities (\ref{equ95b}) into the second static equation
(\ref{equStatic2}) results in the Lane-Emden equation for the total
mass density $\rho^\stat$: 
\be
   \nabla^2 \rho^\stat(r) + k\, \rho^\stat(r) = 0\,,
\fe
which has the well known solution
\be
   \rho^\stat(r) = \rho_0{\sin(r\sqrt{k})\over r\sqrt{k} } \,,
   \quad\text{where}\quad \rho_0 = 1 \,. \mylabel{equ103}
\fe
The requirement that the density vanishes at the surface $r=R$ (where here
\mb{$R=1$}) determines the constant $k$, which is related via
(\ref{equDerivedStruct2}) to the symmetric structure matrix $\S_{\X\Y}$, 
namely
\be
   k = \S_{\X\Y} m^\X m^\Y  = \pi^2 \,. \mylabel{equ99}
\fe
Integrating the density (\ref{equ103}), we obtain the following relation 
between the total mass $M$, radius $R$ and the central density $\rho_0$ 
for this EOS,
\be
   M = {4 \over \pi} \rho_0 R^3 \,.
\fe

\subsection{Analytic solution for the slow--rotation coefficients}
\mylabel{secAnalytic}

Given the ``analytic equation of state'' (\ref{equAnalyticEOS}), the
structure functions are constant, and so are the coefficients in the
differential equations (\ref{equODE0}) and (\ref{equODE2}).  Therefore 
one can write their general {\em regular} solutions as
\bea
    \Phi_0^{\X\Y}(r) &=& {\cal A}^{\X\Y}_0\, {J_{1/2}(r\sqrt{k}) \over 
    \sqrt{r}} + {E^{\X\Y}\over k}\left( r^2 - {6\over k}\right) \nonumber
    \\ 
    &+&{m^\A\S_{\A\B}\C^{\B,\,\X\Y}\over k} \,,  \nonumber \\ 
    \Phi_2^{\X\Y}(r) &=& {\cal A}^{\X\Y}_2\,{J_{5/2}(r \sqrt{k}) \over 
    \sqrt{r}} - {E^{\X\Y}\over k} r^2\,, \mylabel{equAnalyticSolution}
\fea
where $J_{1/2}(x)$ and $J_{5/2}(x)$ are the standard Bessel
functions. From now on we will set $k=\pi^2$ in accordance with the
constraint (\ref{equ99}).  The constants ${\cal A}_0^{\X\Y}$, ${\cal
A}_2^{\X\Y}$, and the $\C^{\A,\,\X\Y}$ are determined by the boundary
conditions  discussed in section \ref{secBC}.  

\subsubsection{The $l = 2$ solution}

The solution for the $l = 2$ component is independent of the chosen 
stellar sequence and is determined by the boundary condition 
(\ref{equBC2}) alone.  This yields
\be
   \Phi^{\X\Y}_2(r) = - {E^{\X\Y} \over \pi^2}\, \left( r^2 - {5 \over 
                      \sqrt{2}} {J_{5/2}(r\pi)\over \sqrt{r}} \right) \,.
                      \mylabel{equAnalytic1}
\fe
Inserting this into the $l=2$ component of (\ref{equRhoXY}) gives the
corresponding total mass density coefficient
\be
   \rho^{\X\Y}_2(r) = - E^{\X\Y} {5 \over \sqrt{2}}\,{J_{5/2}(r\pi) \over 
                      \sqrt{r}} \,,
\fe
while the respective particle density coefficients are found from
(\ref{equNMu2})
\be
   n^{\X\Y}_{\A,\,2} = {k_\A E^{\X\Y} \over \pi^2} \left(r^2 - {5 \over 
   \sqrt{2}} {J_{5/2} (r\pi) \over \sqrt{r}}\right) - E^{\X\Y}_\A r^2 \,.
\fe

\subsubsection{$l = 0$: FCD sequence}

The Fixed Central Density solution for the $l=0$ component is determined by
(\ref{equBC2}) together with (\ref{equFCD}).  In this case one finds that
\be
   \Phi^{\X\Y}_0(r) = {E^{\X\Y} \over \pi^2} \left({3 \sqrt{2} \over 
   \pi^2}\,{J_{1/2} (r\pi) \over \sqrt{r}} + r^2 + {6 \over \pi^2} - 
   3\right) \,.
\fe
From the $l = 0$ component of (\ref{equRhoXY}) we have
\be
   \rho^{\X\Y}_0(r) = - {E^{\X\Y}\over\pi^2} \left( 3\sqrt{2}\,
   {J_{1/2}(r\pi) \over \sqrt{r}} - 6 \right) \,,
\fe
and from (\ref{equNMu2}) it follows that
\bea
   n^{\X\Y}_{\A,\,0} &=& - {k_\A E^{\X\Y} \over \pi^2} \left({3 \sqrt{2} 
   \over\pi^2} {J_{1/2}(r\pi) \over \sqrt{r}} + r^2 - {6 \over \pi^2} 
   \right) \nonumber \\ 
   &+& E^{\X\Y}_\A r^2 \,.
\fea

\subsubsection{$l = 0$: FM sequence}

The Fixed Mass solution for the $l = 0$ component is determined by
(\ref{equBC2}) together with (\ref{equFM}).  One finds
\be
   \Phi^{\X\Y}_0(r) = {E^{\X\Y} \over \pi^2} \left(\sqrt{2}\,
   {J_{1/2}(r\pi) \over \sqrt{r}} + r^2 - 1\right) \,,
\fe
and from the $l = 0$ component of (\ref{equRhoXY}) it follows that
\be
   \rho^{\X\Y}_0(r) = - {E^{\X\Y} \over \pi^2} \left(\pi^2\sqrt{2}
   {J_{1/2}(r\pi) \over \sqrt{r}} - 6\right) \,,
\fe
while (\ref{equNMu2}) leads to
\bea
    n^{\X\Y}_{\A,\,0} &=& - {k_\A E^{\X\Y} \over \pi^2}\left(\sqrt{2} 
    {J_{1/2}(r\pi) \over \sqrt{r}} + r^2 - {6 \over \pi^2} - {3 \over 
    5}\right) \nonumber \\
    &+& E^{\X\Y}_\A\left( r^2 - {3 \over 5}\right) \,.
\mylabel{equAnalyticLast}
\fea

\subsection{Physical parameters of the two-fluid star}
\mylabel{secParams}

The analytic solution (\ref{equAnalytic1})-(\ref{equAnalyticLast}) has 
been expressed entirely in terms of the matrices $E_\A^{\X\Y}$ and 
$E^{\X\Y}$, and we will now discuss the explicit form of these 
coefficients in terms of the physical parameters describing the 
configuration of the two-fluid star. 

The symmetric matrix $\S_{\X\Y}$ has three degrees of freedom. One of
these is already determined by the static radius constraint
(\ref{equ99}) and another is associated with
the $\sigma$ parameter discussed earlier.  A 
further constraint comes from the proton fraction $x_\p$, which we define as
\be
   x_\p \equiv {n_\p^\stat \over n_\n^\stat + n_\p^\stat} \in [0,1] \,,
\fe
and which is constant throughout the static star due to 
(\ref{equProtonFraction}).  In order to simplify expressions, we take the 
proton and neutron masses to be approximately equal, i.e. 
\be
   m^\n \approx m^\p \approx m \,.
\fe
Using (\ref{equ99}) and (\ref{equProtonFraction}), the $k_\X$ are 
expressible as
\be
   k_\n = {\pi^2 \over m} (1 - x_\p) \,, \qaq
   k_\p = {\pi^2 \over m} x_\p \,.
\fe

The ``density structure'' $S_{\X\Y}$ can  be expressed explicitly in 
terms of these parameters as
\be
   \S_{\X\Y} = {\pi^2 \over m^2 (1 + \sigma)} \left(
               \begin{array}{cc}
               \left\{(1 - x_\p) + \sigma(1 - 2 x_\p)\right\} & 
               x_\p \sigma \\
               x_\p \sigma & x_\p \\
               \end{array} \right) \,.
\fe
From this expression we see that, even though \mb{$\sigma\rightarrow
-1$} seemed to be allowed from the ``thermodynamical'' point of view,
it leads to a singular behaviour in the slow-rotation expansion.
Of course, given the fact that the Prakash et al (\cite{PAL88}) 
equation of state indicated that
$\sigma>-0.4$ for all reasonable neutron star densities, cf. Sec.~8.1, 
we do not expect 
this singularity to have any physical significance. 

As we have seen earlier, cf.~(\ref{equRelation1}), the entrainment $\alpha$ 
can be completely described by the effective proton mass $m^{\p*}$, which 
for our EOS (\ref{equAnalyticEOS}) is a constant, and so we can choose 
without loss of generality 
\be
   \beta^\n = 0 \,, \qaq 2 \beta^\p = m \varepsilon \,,
\fe
where we used dimensionless entrainment coefficient $\varepsilon$ defined
earlier in (\ref{equEpsilon}).

With the ``structure functions'' $\S^{\X\Y}$ and $\beta^\X$ expressed
completely in terms of the proton fraction $x_p$, the ``symmetry energy'' 
dependence $\sigma$ and the entrainment coefficient 
$\varepsilon$, the explicit expressions for $E_\A^{\X\Y}$ are
\bea
    E_\n^{\X\Y} &=& {\pi^2 \over 3m (1 + \sigma)} \nonumber \\
                 && \left(\begin{array}{cc} \left\{1 - x_\p + 
                    \sigma(1 - 2 x_\p - x_\p \varepsilon)\right\} & 
                    x_\p \sigma \varepsilon \\
                    x_\p \sigma \varepsilon & x_\p \sigma (1 - 
                    \varepsilon) \\ \end{array}\right) \,,
\label{envar}\fea
and
\be
    E_\p^{\X\Y} = {\pi^2 x_\p \over 3m (1 + \sigma)} 
                  \left(\begin{array}{cc}
                  (\sigma - \varepsilon) & \varepsilon \\
                  \varepsilon & (1 - \varepsilon) \\
                  \end{array}\right) \,,
\label{epvar}\fe
while the coefficient $E^{\X\Y}$ of (\ref{equDerivedStruct2}) is found to be 
\be
   E^{\X\Y} = {\pi^2 \over 3} 
              \left(\begin{array}{cc}
               1 - x_\p (1 + \varepsilon) & x_\p \varepsilon \\
               x_\p \varepsilon & x_\p (1 - \varepsilon) \\
               \end{array}\right) \,. \mylabel{equE}
\fe
It is interesting to note that in the case of co--rotation the
corresponding terms in the analytic solution become
\be
   \Omega_\X E_\A^{\X\Y} \Omega_\Y = {k_\A \over 3 } \Omega^2 \,,
\fe
and 
\be
   \Omega_\X E^{\X\Y} \Omega_\Y = {\pi^2 \over 3} \Omega^2 \,,
\fe
and are therefore seen to be independent of not only the entrainment
$\varepsilon$, but also of $\sigma$. 

\subsection{Explicit results for ellipticity, moments of inertia, and 
Kepler rotation}

Using the analytic solution of Sec.~\ref{secAnalytic} with the physical 
parameters of Sec.~\ref{secParams}, we obtain the following explicit
expression for the ellipticities (\ref{equEllipticity})
\be
e_\A = {3\over2}\left( 1 - {15 \over \pi^2}\right) \Omega_\X E^{\X\Y}
\Omega_\Y - {3\pi^2 \over 2 k_\A} \Omega_\X E_\A^{\X\Y} \Omega_\Y \,,
\fe
where we have set \mb{$R_\A^\stat = 1$}.

For the moments of inertia, 
integrating (\ref{equIA}) for the analytic solution leads to the
explicit result
\be
{\delta I_\A \over I_\A^\stat} = a \, \Omega_\X E^{\X\Y} \Omega_\Y + 
{3\,b \over k_\A} \, \Omega_\X E_\A^{\X\Y} \Omega_\Y\,,
\fe
with coefficients
\be
a = {9\over \pi^2 - 6}
\left( 3 - {\pi^2 \over 5} - {\pi^4 \over 175}\right)\,,
\fe
\be
b = {3 \pi^6 \over 175 (\pi^2 - 6) }\,.
\fe
At the formal level, this result is equivalent to that found in 
earlier work without entrainment (Prix \cite{rp1999}). However, 
the two results  differ at a deeper level since the actual 
matrix elements of $E^{\X\Y}$ and
$E_\A^{\X\Y}$  contain  information about entrainment and the
``symmetry energy'' contribution $\sigma$.

For the analytical solution we can  express the correction to
the Kepler rotation rate (\ref{equdKepler}) as
\be
   \delta\OmKA^2 = {6 \over \pi^2}\left({1 \over 5} - {3 \over \pi^2}
                   \right) \Omega_\X E^{\X\Y} \Omega_\Y - {27 \over 10 
                   k_\A} \Omega_\X E_\A^{\X\Y} \Omega_\Y \,, 
                   \label{analkep}
\fe
where $\A$ denotes the ``outer'' fluid at the equator.

If we take the simple case of the two fluids co-rotating at the maximal
rotation rate, i.e. \mb{$\Omega_\n = \Omega_\p = \OmK$}, this simply leads 
to
\be
   \OmK = {\Omega_\stat \over \sqrt{{6 \over \pi^2} + {3 \over 2}}}
          \approx 0.69 \, \Omega_\stat \approx 0.80\, \sqrt{\pi G 
          \bar{\rho}} \,,
\fe
independent of the parameters of this EOS. 
This is about 20\%--25\% too high compared to the numerical
result \mb{$\OmK\approx0.64\sqrt{\pi G \bar{\rho}}$}. This 
discrepancy originates in the underestimate of the
equatorial radius when approaching $\OmK$, which is illustrated in
Fig.~\ref{lorene}. The ``failure'' of the slow--rotation expansion to
accurately determine the Kepler rotation limit has already been
pointed out by, for example, Salgado et al.~(\cite{SBGH94}).

\section{Exploring the roles of relative rotation, entrainment and 
``symmetry energy''} \label{results}

In this section we will use the analytic solution to the superfluid
slow-rotation problem to
explore the effects of relative rotation, entrainment, and  
``symmetry energy'' on the distribution of matter, the Kepler limit, 
ellipticity, and moments of inertia for a fixed mass star.
The results we present are for a 
particular stellar model with mass $1.4M_\odot$ and radius $10$~km. 
The proton fraction is taken to be $x_\p = 0.1$ and we 
choose the neutron rotation 
rate to be (except in Figs.~\ref{isodens} and \ref{keplerlim}) 
$\Omega_\n = 0.1$. This value corresponds to (in our units) 
the rotation rate of  the fastest
known pulsar (1.6~ms).  
The three parameters that will be varied are $\sigma$, 
$\varepsilon$, and the relative rotation rate
$\Omega_\p/\Omega_\n$.  For the first two parameters
we will consider only the  
values $\sigma = -0.5,0,0.5$ and $\varepsilon = 0,0.4,0.7$, while the 
relative rotation rate will be assumed to lie in the range  $-2 \leq 
\Omega_\p/\Omega_\n \leq 2$ (which allows the protons to be 
counter-rotating with respect to the neutrons).  The values chosen for 
$\sigma$ are in accordance with the discussion in 
Sec.~\ref{analeos}.  We also recall that the best current
estimates for entrainment 
imply a range of $0.4 \leq \varepsilon \leq 0.7$. This means that
our chosen models correspond to the 
expected upper and lower limits, as well as the no-entrainment limit (which
provides a useful reference).  

\subsection{Effects on the ``local'' structure}

First we focus on the ``local'' structure by examining how the 
particles get redistributed throughout the star because of rotation.  In 
Fig.~\ref{densneuts} we show plots of the rotationally-induced corrections 
to the neutron number densities in the equatorial plane and along the 
polar axis.  We show plots for $\sigma = 0$ only because other values 
lead to similar results.  An inset is provided to show that the 
effect of entrainment is small but not completely negligible.  Recall that 
we are considering the fixed mass solution, so the density at the center 
of the star will decrease as the star is spun up.  This also explains why the 
density correction is negative along the polar axis. 
Furthermore,  the density 
corrections in the equatorial plane are positive near the stellar 
surface since 
rotation causes the matter to bulge.  Fig.~\ref{densprots} 
shows the corresponding proton particle number density 
corrections.  Unlike in the case for the neutrons, we have added a
graph for $\sigma = 0.5$. From Fig.~\ref{densprots}  
we notice that the effect of changing $\sigma$ and $\varepsilon$ 
 is more pronounced for the protons.  
That this should be the case is easy to understand since the  proton 
fraction is small and only $10\%$ of the mass of the star is in the 
protons. The relative effect of a changing $\varepsilon$ is more 
apparent in the 
proton plots, then, because the absolute magnitude of the proton number 
density corrections are always $10\%$ of those of the neutrons.
From (\ref{equAnalyticLast}),  ({\ref{envar}) and (\ref{epvar}) we see
that the main difference between the two cases is that while 
changes in $\varepsilon$ and $\sigma$ affect $E_\p^{\X\Y}$ at leading order 
($\sim x_p$), they only lead to higher order corrections to $E_\n^{\X\Y}$
(since there are also terms of $O(1)$ in (\ref{envar})).
We can also see from the right-hand panel of 
Fig.~\ref{densprots} that changing 
$\sigma$ from $0$ to $0.5$ results in substantial modifications.
This shows clearly that the ``symmetry energy'' 
can play an important role in determining the 
rotational configuration of a two-fluid star.

\begin{figure}
   \resizebox{\hsize}{!}{\includegraphics[clip]{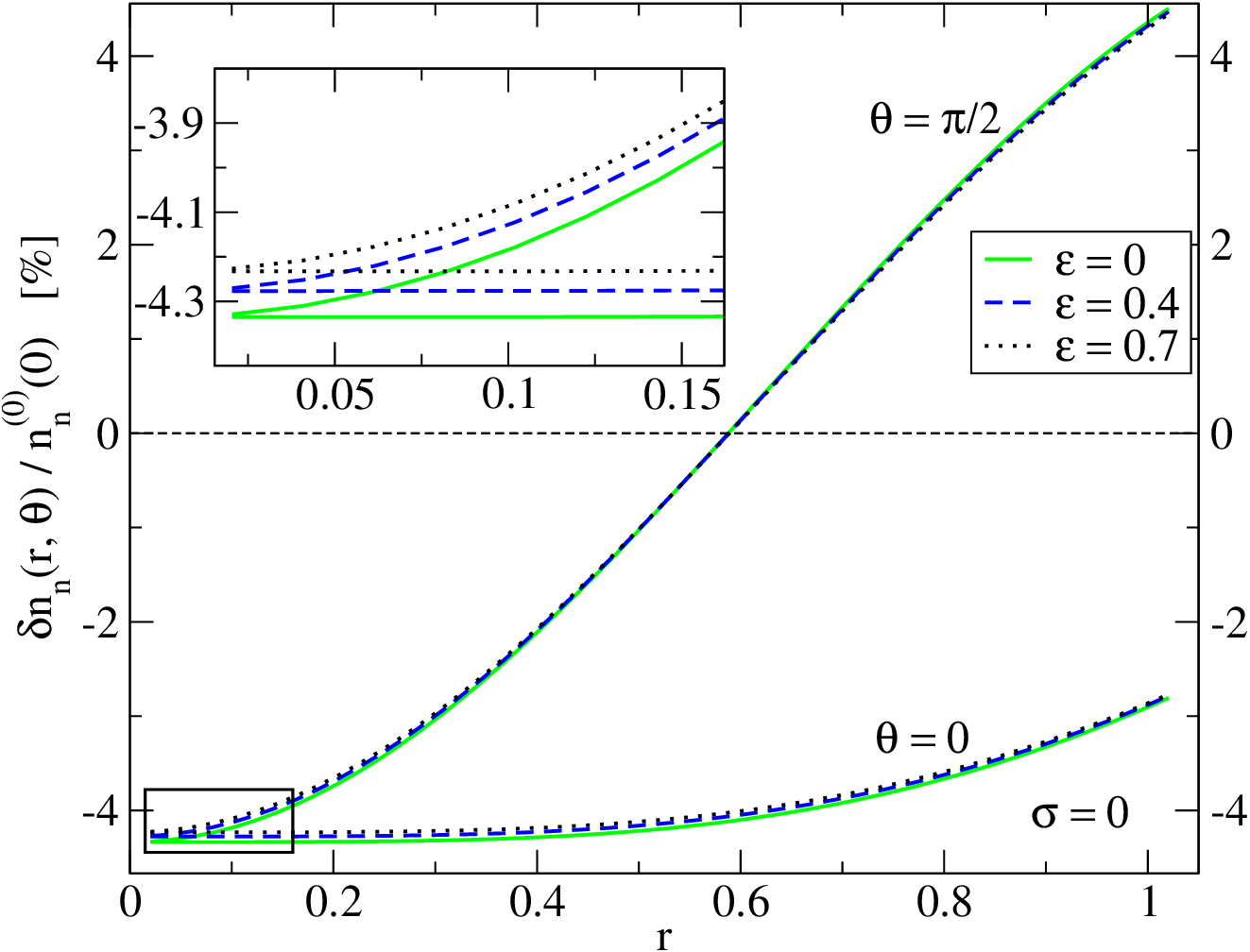}} 
   \caption{Plots of the radial profiles of  the 
   neutron density corrections, normalized by the static number density, 
	in  the equatorial plane 
   ($\theta = \pi/2$) and along the polar axis ($\theta = 0$)
   for $\sigma = 0$, $\varepsilon = 0,0.4,0.7$, and $\Omega_\n = 0.1$ and 
   $\Omega_\p = 0.25 \Omega_\n$.  In this, and the following figures, the 
   dotted lines correspond to $\varepsilon = 0.7$, dashed have $\varepsilon = 
   0.4$, whereas the solid lines are for $\varepsilon = 0$.}
   \label{densneuts}
\end{figure}   

\begin{figure*}
   \centering
   \includegraphics[height = 6cm,clip]{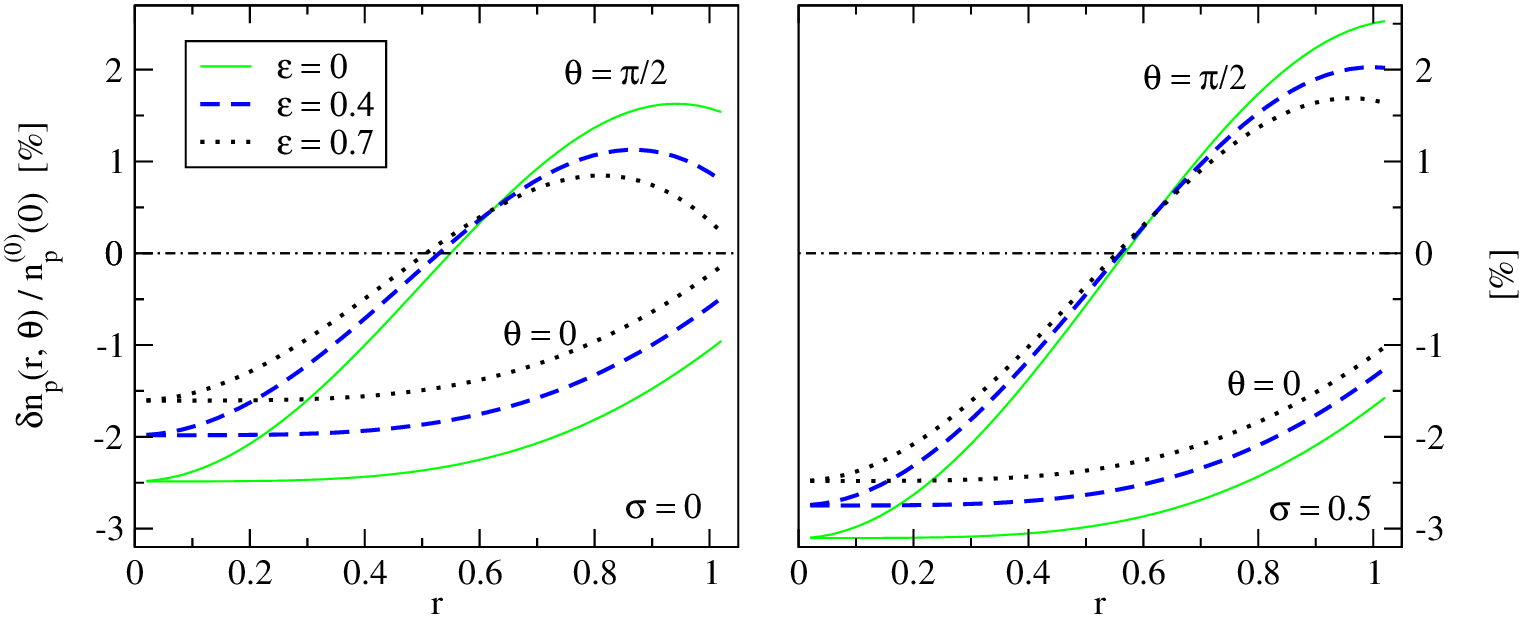} 
   \caption{Plots of the radial profiles of the 
   proton density corrections, normalized by the static number density,  
in the equatorial plane 
   ($\theta = \pi/2$) and along the polar axis ($\theta = 0$)
   for $\sigma = 0,0.5$, $\varepsilon = 0,0.4,0.7$, and $\Omega_\n = 0.1$ 
   and $\Omega_\p = 0.25 \Omega_\n$.}
   \label{densprots}
\end{figure*} 

Fig.~\ref{isodens} provides a different perspective on the 
rotationally-induced redistribution of the particles. Here we 
show isodensity surfaces in the $(r,\theta)$-plane, for $\varepsilon = 0$, 
$\sigma = 0$ (in the left panel) and $\varepsilon = 0.7$, $\sigma = - 0.5$ 
(in the right panel).  Note that we have considered slightly larger values 
of the rotation rates (just below the Kepler limit) 
in order to exaggerate certain effects.  For a given 
density, we compare the isodensity curves for a non-rotating star 
to the neutron and proton isodensity curves for the rotating model.  In both 
panels we see that the neutron and proton curves actually intersect each 
other near the surface of the star.  That is, along the equator the 
neutrons actually extend further than the protons, while  the 
protons  extend further along the rotation axis. 
This means that, the rotational configuration of the protons is, in fact,
prolate. We note that this 
effect has been exaggerated in the panels, because of the higher rotation 
rates, but it happens also for lower rotation rates
(as used in the earlier 
figures).  Near the center of each panel, we see that the neutron and 
proton surfaces no longer intersect.  This is no doubt due to the fact 
that the centrifugal forces are smaller closer to the center of the 
star.  

\begin{figure*}
   \centering
   \includegraphics[height=7cm,clip]{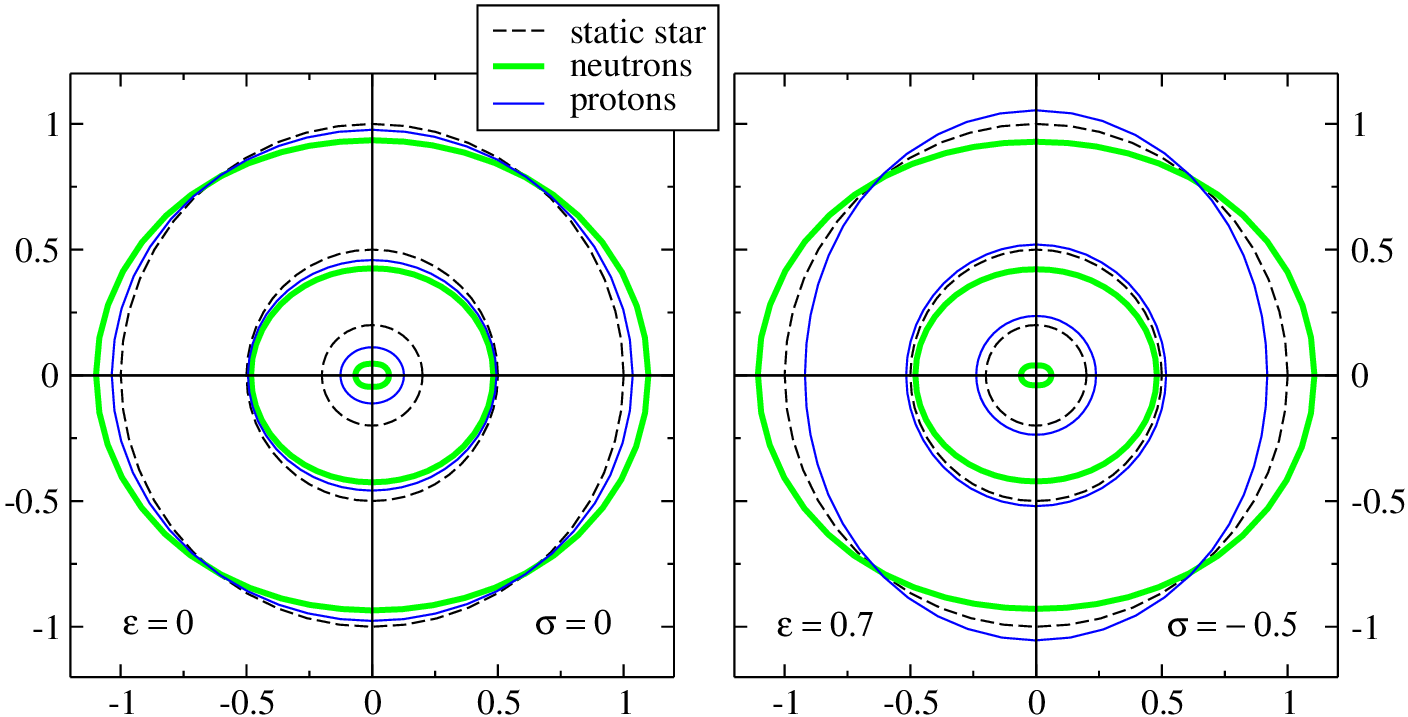} 
   \caption{Neutron (thick line), proton (thin line) and static 
   configuration (dashed line) isodensity curves in the meridional plane,   
   for $\sigma = 0,- 0.5$, $\varepsilon = 0,0.7$, and $\Omega_\n = 0.15$ 
   and $\Omega_\p = 0.25 \Omega_\n$.}
   \label{isodens}
\end{figure*}

\subsection{Effects on the ``global'' structure}

Now we focus our attention on the roles of $\sigma$, $\varepsilon$, 
and the relative rotation rate $\Omega_\p/\Omega_\n$ in determining 
macroscopic properties of the star; in particular, the proton and neutron 
ellipticities, their respective moments of inertia, and the Kepler limit.  
As before, we will keep the mass of the star fixed.  

Fig.~\ref{ellps} illustrates the neutron and proton ellipticities as 
functions of the relative rotation rate, and Fig.~\ref{mominert} gives 
their moments of inertia (normalized by their static values).  There is 
an obvious quadratic behavior in each plot due simply to the slow-rotation 
expansion.  As well, the intersection of all curves at 
$\Omega_\p/\Omega_\n = 1$ occurs because the protons then co-rotate with the 
neutrons and the system is behaving as a single fluid.  
Notice that entrainment has the largest influence when the neutrons and 
protons counter-rotate.  This is easily understood as a consequence of the 
basic fact that the entrainment parameter
represents the way that the equation of state (as represented by the energy 
functional) depends on $|\vec{v}_\n - \vec{v}_\p|^2$.  
The protons are in general much more affected by changes in the various
parameters than the neutrons,  again due to the 
fact that the neutrons carry the bulk of the mass of the star.  Perhaps 
most interesting is the effect of both the entrainment parameter
$\varepsilon$ and ``symmetry energy'' parameter $\sigma$ in
determining the minima of  the curves.  For both the protons and the
neutrons, we see that an increase  in $\varepsilon$ for a fixed value
of $\sigma$ leads to a deeper value for the minimum. Decreasing the
value of $\sigma$  causes the minima to become even deeper.  In
particular, we see from the left-most panel in Fig.~\ref{ellps} that
the minima have become deep enough that the protons can be prolate
(i.e. have negative ellipticity) even though they rotate in the same
direction as the neutrons.  Finally, we note that as $\sigma$ is
decreased, the neutron ellipticities go from having minima in the
right-most and center panels, to having maxima in the left-most panel.
That is, if the absolute value of the relative rotation could be made
large enough the neutron fluid could also become prolate.    

It is interesting to compare the curves for the ellipticities in
Fig.~\ref{ellps} and the corresponding moments of inertia in
Fig.~\ref{mominert}, for example for \mb{$\sigma=0,\,\varepsilon=0.4$}
(middle panel, dashed line) for the protons. For \mb{$\Omega_\p=0$} we
observe that the proton fluid is nearly spherical
(\mb{$e_\p\approx0$}), and still its moment of inertia is {\em higher}
than of the static (spherical) configuration $\delta I_\p>0$, and this
is simply because the mass distribution has been shifted further away
from the rotation axis even though the surface itself is (nearly)
unchanged in this case. This can also be clearly seen in the density
correction in Fig.~\ref{densprots}, for example for
\mb{$\sigma=0,\,\varepsilon=0.7$} and \mb{$\Omega_\p/\Omega_\n=0.25$} (left
panel, dotted line), which nearly vanishes at the surface $r=1$, and
is to be compared to the  corresponding global quantities in
Fig.~\ref{ellps} and Fig.~\ref{mominert}. 

\begin{figure*}
   \centering
   \includegraphics[clip,width = 17cm]{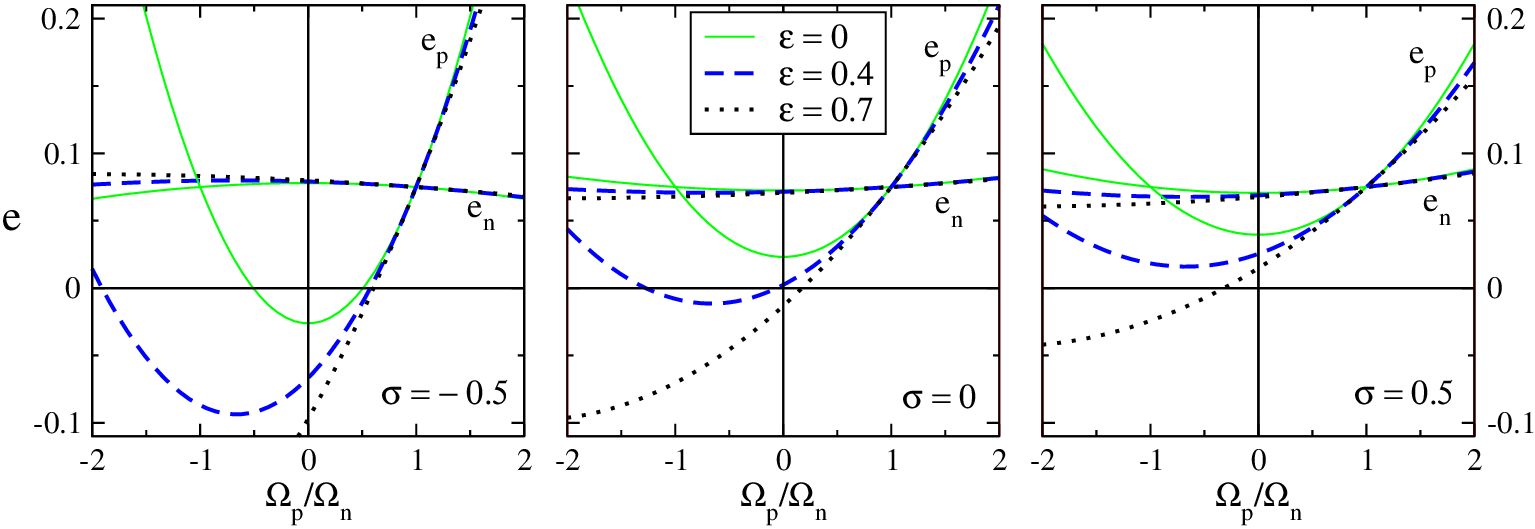} 
   \caption{The neutron and proton ellipticities as functions of the 
   relative rotation $\Omega_\p/\Omega_\n$, for $\sigma = -0.5,0,0.5$, 
   $\varepsilon = 0,0.4,0.7$, and $\Omega_\n = 0.1$.}
   \label{ellps}
\end{figure*}

\begin{figure*}
   \centering
   \includegraphics[clip,width = 17cm]{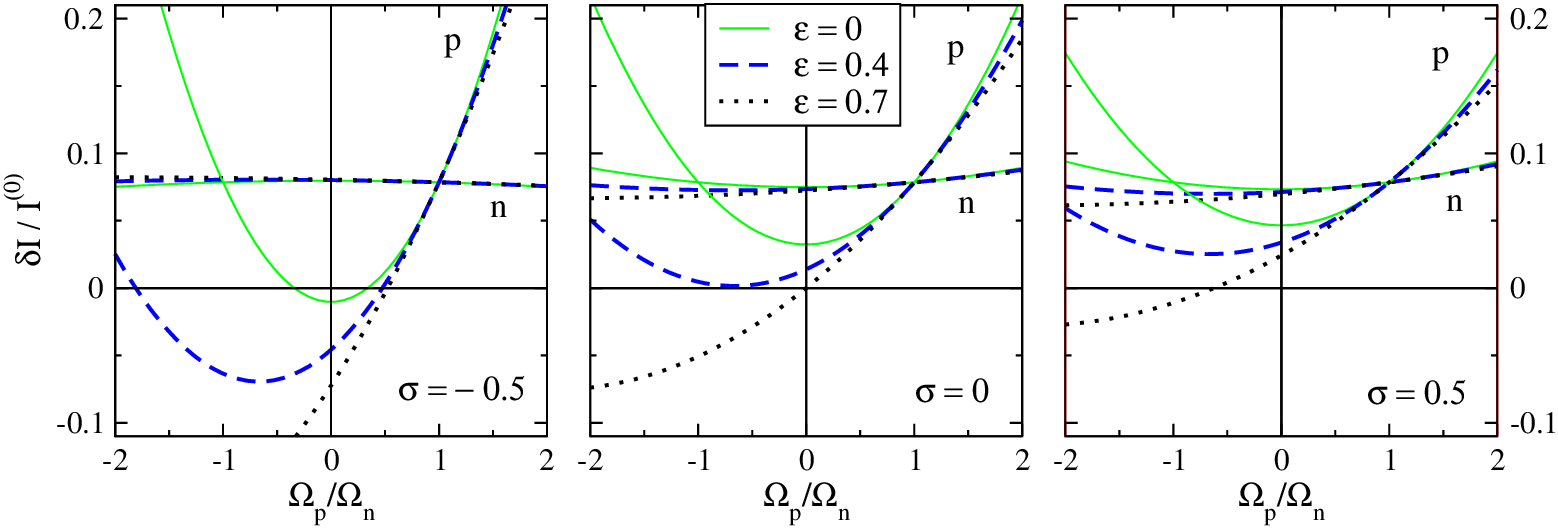}
   \caption{Neutron ($\n$ ) and proton ($\p$ ) 
moments of 
   inertia for $\sigma = -0.5,0,0.5$, $\varepsilon = 0,0.4,0.7$, and 
   $\Omega_\n = 0.1$.}
   \label{mominert}
\end{figure*}

Finally, results for the Kepler, or mass-shedding, limit are shown in 
Fig.~\ref{keplerlim}.  To understand these results one must appreciate 
that there is a subtle difference with 
the single-fluid case: In our case the two fluids  can rotate 
independently at different rates. Thus, one of the fluids 
typically extends beyond 
the other, in particular  at the equator.  Since the Kepler limit is 
defined by the outermost fluid at the equator, we can use Eq. 
(\ref{fullkep}) in the following way: When the neutrons are outermost, set 
$A = \n$ and $\Omega_\n = \Omega_{K,\n}$ and solve the resulting 
quadratic for $\Omega_{K,\n}$ as a function of the ratio 
$\Omega_\n/\Omega_\p$, and vice versa in the case when the 
protons extend beyond the neutrons.  In 
Fig.~\ref{keplerlim} we  show the resultant solutions over the 
entire range of the relative rotation rate.  The Kepler rate is easy to 
determine, however, because it is given by the neutron curves when 
$\Omega_\n/\Omega_\p \ge 1$, and the proton curves when 
$\Omega_\n/\Omega_\p \le 1$.  Of course, the various curves always intersect 
when $\Omega_\n/\Omega_\p = 1$,  the case that corresponds 
to corotation of the two fluids.  For the 
case of $\sigma = \epsilon = 0$, we find results in good qualitative 
agreement with the relativistic study of Andersson \& Comer (\cite{AC01a}).  
In particular, 
we see that the Kepler limit changes little when $\Omega_\n \ge 
\Omega_\p$.  As explained by Andersson \& Comer (\cite{AC01a}), this
is due to the fact that the neutrons make up $90\%$ of the mass of the
star, and the star is behaving like a single-fluid star with a small proton 
component.  When $\Omega_\p \ge \Omega_\n$ the 
Kepler limit increases as $\Omega_\n$ is decreased.  Again, this is 
natural because  the neutrons still dominate the mass of the star, and the 
Kepler rate is simply approaching the non-rotating star limit.  
    
\begin{figure*}
   \centering
   \includegraphics[clip,width = 17cm]{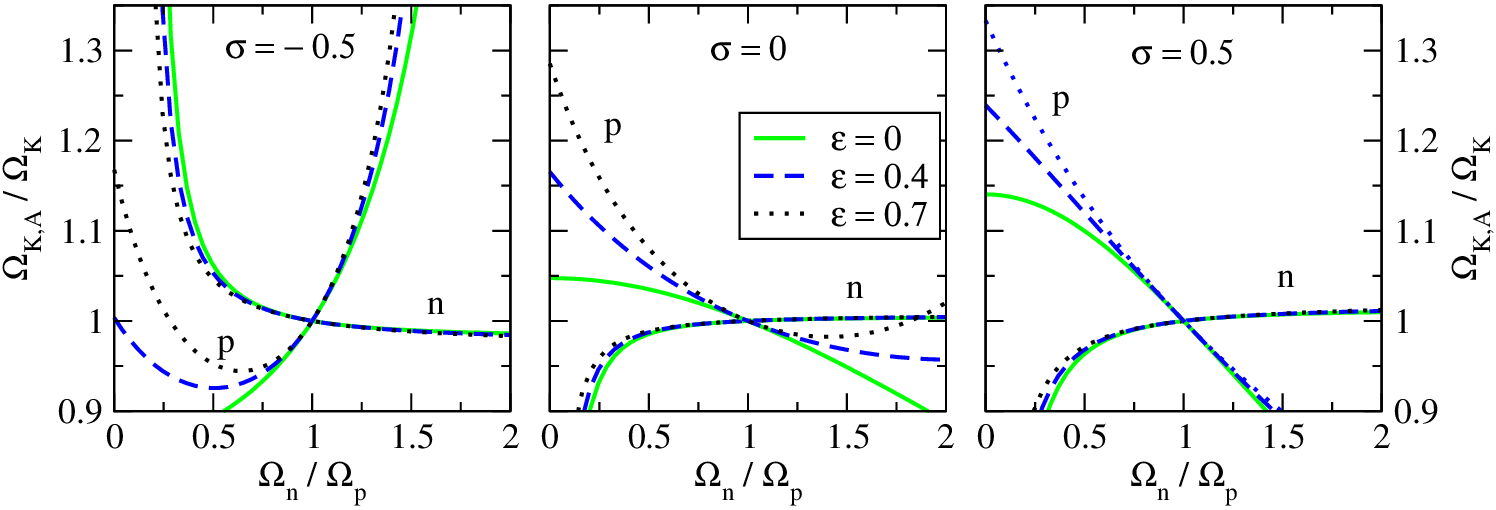} 
   \caption{Plots of the neutron (n) and proton (p) Kepler limits as functions 
   of the relative rotation $\Omega_\n/\Omega_\p$, for 
   $\sigma = -0.5,0,0.5$ and $\varepsilon = 0,0.4,0.7$.}
   \label{keplerlim}
\end{figure*}

\section{Conclusions}

We have developed a formalism for modeling slowly-rotating 
Newtonian superfluid neutron stars incorporating entrainment.  
We have used a two-fluid description, where one fluid is the 
superfluid neutrons and the other 
is a charge-neutral conglomerate of the remaining constituents.
A detailed discussion of the relation between entrainment 
and nuclear physics calculations (i.e.~equations of state) was given.  
Using an equation of state that is quadratic in both the mass-densities 
and relative velocities of the fluids, we found that an analytic solution 
to the slow-rotation equations could be obtained.  
This solution is the natural 
extension to the two-fluid case of the \mb{$\energy\propto\rho^2$}
polytrope in the single fluid case (which has proven to be very
useful for understanding the  properties of ordinary fluid neutron
stars).  We used the analytic solution to explore effects due to relative 
rotation,
entrainment, and  ``symmetry energy'' on both the ``local'' and
``global'' properties of a fixed-mass star.  An unexpected result
was that the ``symmetry   energy'' parameter had as much impact on the
rotational equilibria as the  entrainment parameter.  

Our ultimate goal is to study the modes of oscillation of both Newtonian 
and general relativistic slowly rotating superfluid neutron stars.  We 
believe that the formalism and analytic solution presented here will be 
valuable in reaching this goal.  In particular, the inclusion of 
entrainment is absolutely necessary in determining how the dominant 
dissipative mechanism (the so-called mutual friction) of rotating 
superfluid neutron stars affects the gravitational radiation emitted from 
unstable modes.

\begin{acknowledgements}
We would like to thank the members of the Observatory of Paris-Meudon 
Numerical Relativity group for providing us with their LORENE code, and 
J.~Novak for fruitful conversations.  NA is Leverhulme Prize fellow
and acknowledges support 
from PPARC via grant number PPA/G/S/1998/00606.

RP and NA acknowledge support from the EU Programme 'Improving the Human
Research Potential and the Socio-Economic Knowledge Base' (Research
Training Network Contract HPRN-CT-2000-00137).

GC gratefully acknowledges partial support from a Saint Louis
University SLU2000 Faculty Research Leave Award as well as EPSRC in
the UK via grant number GR/R52169/01 (to NA), and the warm hospitality of the
Center for Gravitation and Cosmology of the University of Wisconsin-Milwaukee 
and the University of Southampton where part of this research was carried out.   
\end{acknowledgements}

\appendix

\section{An alternative derivation of the first integrals of motion}
\mylabel{secIntegrals}

A more elegant way of finding the first integrals of motions
(\ref{equIntegrals}) consists of using
the identity     
\be
v^\b\nabla_\b \,p_\a = \Lie_v p_\a - p_\b \nabla_\a v^\b\,,
\fe
in terms of the Lie derivative $\Lie_v$, which allows one to rewrite
Euler's equations (\ref{equGenEuler}) in the form 
\bea
\partial_t p^\n_\a &+& \Lie_{v_\n} p^\n_\a - \nabla_\a \left(p^\n_0+
v_\n^\b p^\n_\b \right) = 0\,, \\
\partial_t p^\p_\a &+& \Lie_{v_\p} p^\p_\a - \nabla_\a \left(p^\p_0 +
v_\p^\b p^\p_\b \right) = 0\,.
\fea
Stationarity (\mb{$\partial_t p_\a^\X=0$}), uniform rotation
(\mb{$v_\X^\a = \Om_\X\varphi^\a$}) and axial symmetry
(\mb{$\Lie_\varphi p_\a^\X=0$}) then directly results in the integrals
\bea
\nabla_\a \left(p^\n_0 + v_\n^\b p^\n_\b \right) &=& 0\,, \\
\nabla_\a \left(p^\p_0 + v_\p^\b p^\p_\b \right) &=& 0\,.
\fea
and inserting (\ref{equp0}) this explicitly becomes
\bea
\mu^\n + m^\n \Phi - {1\over2}m^\n v_\n^{\,2}  &=& \C^\n\,, \\
\mu^\p + m^\p \Phi - {1\over2}m^\p v_\p^{\,2}  &=& \C^\p\,.
\fea

\end{document}